\newcommand\apjcls{1}
\newcommand\aastexcls{2}
\newcommand\othercls{3}

\newcommand\papercls{\aastexcls}
\documentclass[tighten, times, twocolumn]{aastex62} 

\if\papercls \apjcls
\usepackage{apjfonts}
\else\if\papercls \othercls
\usepackage{epsfig}
\usepackage{margin}
\usepackage{times}
\fi\fi
\usepackage{ifthen}
\usepackage{natbib}
\usepackage{amssymb, amsmath}
\usepackage{appendix}
\usepackage{etoolbox}
\usepackage[T1]{fontenc}
\usepackage{paralist}

\if\papercls \apjcls
\newcommand\aas{\ref@jnl{AAS Meeting Abstracts}}
\newcommand\dps{\ref@jnl{AAS/DPS Meeting Abstracts}}
\newcommand\maps{\ref@jnl{MAPS}}
\else\if\papercls \othercls
\usepackage{astjnlabbrev-jh}
\fi\fi

\if\papercls \aastexcls
\hypersetup{citecolor=blue, 
            linkcolor=blue, 
            menucolor=blue, 
            urlcolor=blue}  
\else
\usepackage[
bookmarks=true,           
bookmarksnumbered=true,   
colorlinks=true,          
citecolor=blue,           
linkcolor=blue,           
menucolor=blue,           
urlcolor=blue,            
linkbordercolor={0 0 1},  
pdfborder={0 0 1},
frenchlinks=true]{hyperref}
\fi

\if\papercls \othercls

\else

\fi

\providecommand{\adsurl}[1]{\href{#1}{ADS}}

\makeatletter
\patchcmd{\NAT@citex}
  {\@citea\NAT@hyper@{%
     \NAT@nmfmt{\NAT@nm}%
     \hyper@natlinkbreak{\NAT@aysep\NAT@spacechar}{\@citeb\@extra@b@citeb}%
     \NAT@date}}
  {\@citea\NAT@nmfmt{\NAT@nm}%
   \NAT@aysep\NAT@spacechar\NAT@hyper@{\NAT@date}}{}{}

\patchcmd{\NAT@citex}
  {\@citea\NAT@hyper@{%
     \NAT@nmfmt{\NAT@nm}%
     \hyper@natlinkbreak{\NAT@spacechar\NAT@@open\if*#1*\else#1\NAT@spacechar\fi}%
       {\@citeb\@extra@b@citeb}%
     \NAT@date}}
  {\@citea\NAT@nmfmt{\NAT@nm}%
   \NAT@spacechar\NAT@@open\if*#1*\else#1\NAT@spacechar\fi\NAT@hyper@{\NAT@date}}
  {}{}
\makeatother

\makeatletter
\DeclareRobustCommand{\lowcase}[1]{\@lowcase#1\@nil}
\def\@lowcase#1\@nil{\if\relax#1\relax\else\MakeLowercase{#1}\fi}
\makeatother

\DeclareSymbolFont{UPM}{U}{eur}{m}{n}
\DeclareMathSymbol{\umu}{0}{UPM}{"16}
\let\oldumu=\umu
\renewcommand\umu{\ifmmode\oldumu\else\math{\oldumu}\fi}

\if\papercls \othercls

\else

\fi

\let\oldsim=\sim
\renewcommand\sim{\ifmmode\oldsim\else\math{\oldsim}\fi}
\let\oldpm=\pm
\renewcommand\pm{\ifmmode\oldpm\else\math{\oldpm}\fi}
\newcommand\by{\ifmmode\times\else\math{\times}\fi}

\newbox{\wdbox}
\renewcommand\c{\setbox\wdbox=\hbox{,}\hspace{\wd\wdbox}}
\renewcommand\i{\setbox\wdbox=\hbox{i}\hspace{\wd\wdbox}}

\newcount\timect
\newcount\hourct
\newcount\minct
\newcommand\now{\timect=\time \divide\timect by 60
         \hourct=\timect \multiply\hourct by 60
         \minct=\time \advance\minct by -\hourct
         \number\timect:\ifnum \minct < 10 0\fi\number\minct}

\catcode`@=11

\newcommand\comment[1]{}

\newcommand\commenton{\catcode`\%=14}

\renewcommand\math[1]{$#1$}
\newcommand\mathshifton{\catcode`\$=3}

\let\atab=&
\newcommand\atabon{\catcode`\&=4}

\let\oldmsp=\sp
\let\oldmsb=\sb
\def\sp#1{\ifmmode
           \oldmsp{#1}%
         \else\strut\raise.85ex\hbox{\scriptsize #1}\fi}
\def\sb#1{\ifmmode
           \oldmsb{#1}%
         \else\strut\raise-.54ex\hbox{\scriptsize #1}\fi}
\newbox\@sp
\newbox\@sb
\def\sbp#1#2{\ifmmode%
           \oldmsb{#1}\oldmsp{#2}%
         \else
           \setbox\@sb=\hbox{\sb{#1}}%
           \setbox\@sp=\hbox{\sp{#2}}%
           \rlap{\copy\@sb}\copy\@sp
           \ifdim \wd\@sb >\wd\@sp
             \hskip -\wd\@sp \hskip \wd\@sb
           \fi
        \fi}
\def\msp#1{\ifmmode
           \oldmsp{#1}
         \else \math{\oldmsp{#1}}\fi}
\def\msb#1{\ifmmode
           \oldmsb{#1}
         \else \math{\oldmsb{#1}}\fi}

\def\supon{\catcode`\^=7}

\def\subon{\catcode`\_=8}

\def\supsubon{\supon \subon}

\newcommand\actcharon{\catcode`\~=13}

\newcommand\paramon{\catcode`\#=6}

\comment{And now to turn us totally on and off...}

\newcommand\reservedcharson{ \commenton  \mathshifton  \atabon  \supsubon 
                             \actcharon  \paramon}

\catcode`@=12
\reservedcharson

\if\papercls \apjcls

\else

\fi

\newcommand\chisq{\ifmmode{\chi\sp{2}}\else\math{\chi\sp{2}}\fi}
\newcommand\redchisq{\ifmmode{ \chi\sp{2}\sb{\rm red}}
                    \else\math{\chi\sp{2}\sb{\rm red}}\fi}
\newcommand\Teq{\ifmmode{T\sb{\rm eq}}\else$T$\sb{eq}\fi}
\newcommand\mjup{\ifmmode{M\sb{\rm Jup}}\else$M$\sb{Jup}\fi}
\newcommand\rjup{\ifmmode{R\sb{\rm Jup}}\else$R$\sb{Jup}\fi}
\newcommand\msun{\ifmmode{M\sb{\odot}}\else$M\sb{\odot}$\fi}
\newcommand\rsun{\ifmmode{R\sb{\odot}}\else$R\sb{\odot}$\fi}
\newcommand\mearth{\ifmmode{M\sb{\oplus}}\else$M\sb{\oplus}$\fi}
\newcommand\rearth{\ifmmode{R\sb{\oplus}}\else$R\sb{\oplus}$\fi}

\usepackage{CJKutf8} 
\usepackage{natbib}
\usepackage{xcolor}
\usepackage[para,online,flushleft]{threeparttable}
\usepackage{booktabs}
\usepackage{amsmath}

\shortauthors{Zaritsky, et al.}
\shorttitle{}

\begin{document}

\title{Systematically Measuring Ultra Diffuse Galaxies (SMUDGes). I.  Survey Description and First Results in the Coma Galaxy Cluster and Environs}
  
\author{Dennis Zaritsky}
\affiliation{Steward Observatory and Department of Astronomy, University of Arizona, 933 N. Cherry Ave., Tucson, AZ 85721, USA; dennis.zaritsky@gmail.com}

\author{Richard Donnerstein}
\affiliation{Steward Observatory and Department of Astronomy, University of Arizona, 933 N. Cherry Ave., Tucson, AZ 85721, USA}

\author{Arjun Dey}
\affiliation{NOAO, National Optical Astronomy Observatory, 950 N. Cherry Ave., Tucson, AZ 85719, USA}

\author{Jennifer Kadowaki}
\affiliation{Steward Observatory and Department of Astronomy, University of Arizona, 933 N. Cherry Ave., Tucson, AZ 85721, USA}

\author{Huanian Zhang\begin{CJK*}{UTF8}{gkai}(张华年)\end{CJK*}}
\affiliation{Steward Observatory and Department of Astronomy, University of Arizona, 933 N. Cherry Ave., Tucson, AZ 85721, USA}

\author{Ananthan Karunakaran}
\affiliation{Department of Physics, Engineering Physics and Astronomy Queen's University Kingston, ON K7L 3N6, Canada}

\author{David Mart\'inez-Delgado}
\affiliation{Astronomisches Rechen-Institut, Zentrum für Astronomie der Universität Heidelberg, Mönchhofstr. 12-14, D-69120 Heidelberg, Germany}
\affiliation{Max-Planck-Institut für Astronomie, Königstuhl 17, D-69117 Heidelberg, Germany}

\author{Mubdi Rahman}
\affiliation{Dunlap Institute for Astronomy and Astrophysics, University of Toronto, 50 St. George Street, Toronto, Ontario M5S 3H4, Canada}

\author{Kristine Spekkens}
\affiliation{Department of Physics, Engineering Physics and Astronomy Queen's University Kingston, ON K7L 3N6, Canada}
\affiliation{Department of Physics and Space Science Royal Military College of Canada P.O. Box 17000, Station Forces Kingston, ON K7K 7B4, Canada}

\email{dennis.zaritsky@gmail.com}


\begin{abstract}
We present a homogeneous catalog of 275 large (effective radius $\gtrsim$ 5.3\arcsec) ultra-diffuse galaxy (UDG) candidates lying within an $\approx$ 290 square degree region surrounding the Coma cluster. The catalog results from our automated postprocessing of data from the Legacy Surveys, a three-band imaging survey covering 14,000 square degrees of the extragalactic sky.
We describe a pipeline that identifies UDGs and provides their basic parameters. The survey is as complete in these large UDGs as previously published UDG surveys of the central region of the Coma cluster. We conclude that the majority of our detections are at roughly the distance of the Coma cluster, implying effective radii $\ge 2.5$ kpc, and that our sample contains a significant number of analogs of DF 44, where the effective radius exceeds 4 kpc, both within the cluster and in the surrounding field. 
The $g-z$ color of our UDGs spans a large range, suggesting that even  large UDGs may reflect a range of formation histories. A majority of the UDGs are consistent with being lower stellar mass analogs of red sequence galaxies, but we find both red and blue UDG candidates in the vicinity of the Coma cluster and a relative overabundance of blue UDG candidates in the lower density environments and the field. Our eventual processing of the full Legacy Surveys data will produce the largest, most homogeneous sample of large UDGs. 
\end{abstract}


\section{Introduction}
\label{sec:intro}

To test the dark matter paradigm and provide boundary conditions for theories of galaxy formation, 
the astronomical community has been striving to measure the mass-to-light ratio (M/L) of galaxies for over 40 years \citep[e.g.][]{roberts,rubin}. As of a couple of years ago, the developed consensus was that M/L within the luminous portion of galaxies increases dramatically only when one considers extremely low mass galaxies, such as the dwarf spheroidal and ultra-faint satellites of the Milky Way for which M/L reaches values of several hundreds or even above a thousand in solar units \citep{wolf}.  That consensus has recently been challenged by the discovery of apparently massive, low surface brightness, dark matter dominated galaxies \citep{vdk15a}, broadly referred to as ultra diffuse galaxies (UDGs).

Large galaxies of low surface brightness that elude standard galaxy catalogs have been a focus of study for decades \citep{disney,sandage,impey,schombert,sch,sprayberry,dalcanton}. Such work had even occasionally highlighted physically large low surface brightness galaxies such as Malin 1 \citep{impey89}.
Recently, ever more sensitive observations \citep[cf.][]{koda,mihos,munoz,lee,roman17a,shi,vdb17,venhola,wittman,greco} have helped investigators reach even fainter surface brightnesses and, in the case of the Coma galaxy cluster survey by \cite{vdk15a}, identify many such galaxies that are large and quiescent. Because the galaxies in that particular study are clustered on the sky about the Coma cluster, distance-by-association allowed those investigators to convert angular sizes to physical ones. This aspect is critical because redshift measurements become increasingly difficult as we focus our attention on galaxies with ever lower surface brightnesses. Spectroscopic redshifts exist for only a handful of these galaxies and, for the most part, confirm the estimated distances \citep{vdk15b,kadowaki17,alabi}. With  physical sizes in hand, \cite{vdk15a} found that their UDGs are surprisingly large, including many systems with effective radii, $r_e$, as large as that of the Milky Way. The survival of such systems in the cluster tidal field suggests that these objects are dark matter dominated. Analogous systems, of somewhat brighter surface brightness and smaller size, had previously been found in the Perseus cluster \citep{penny} and the tidal argument was used there as well to infer large dark matter content.

To demonstrate that the large size of some of the UDGs connotes correspondingly large mass requires independent mass estimates. In the case of one Coma UDG, DF 44, a stellar velocity dispersion was subsequently measured \citep{vdk16}. The velocity dispersion is indeed consistent with that expected if that galaxy lies in a massive dark matter halo, but the measurement is taken at small radius relative to the calculated virial radius and so weakly constrains the total halo mass. Furthermore, the exposure time required to obtain this measurement was extreme and analogous measurements now exist for only a couple of other UDGs \citep{beasleya,vdk17}. 
Exploiting the relationship between the number of globular clusters and total galaxy mass established for normal galaxies \citep{blakeslee,georgiev,harris,forbes,harris17,zar17b} provides an alternate avenue for estimating the total masses of larger samples \citep[cf.][]{amorisco}. The large populations of globular clusters found in some UDGs
support the conjecture that those UDGs lie in massive ($M > 10^{11} M_\odot$) dark matter halos \citep{vdk17}. Lastly, the application of galaxy scaling relations also supports the conjecture that the largest UDGs are indeed massive \citep{zar17a}.

As these intriguing results developed, other studies have concluded that UDGs are predominantly low mass galaxies \citep{amorisco17,sifon}. As in any astronomical population, the small, low mass objects dominate by number over the large, massive ones. This phenomenon is clear in the comparison of the sizes and luminosities of the \cite{vdk15a} and \cite{yagi} Coma cluster UDG samples. However, the details of the UDG mass distribution are critical in discriminating between the variety of formation models that have already been developed, which invoke a range of physical phenomena such as environmental quenching, feedback, and high specific angular momenta, to explain both the low apparent star formation efficiencies and large sizes of UDGs \citep{yozin,amorisco,agertz,dic,chan,carleton}. Arguments about the nature of UDGs and their formation history should bear in mind that UDGs, being selected solely by surface brightness, are likely to include diverse objects and do not have a single origin story \citep{zar17a,ferre,lim}.

This large range of properties among UDGs and the extensive literature history of low surface brightness galaxies already cited raise the question of whether the term UDG is simply a renaming of an already identified class of galaxy, such as the Low Mass Cluster Galaxy (LMCG) class presented by \cite{conselice}. It is early in the detailed internal study of these objects, but the physically largest of these, with halo masses that are inferred to be close to that of an L$^*$ galaxy and globular cluster populations to match, appear to be a new category of object, or at least represent a new emphasis. Unfortunately, the current UDG definition, based on central surface brightness, $\mu_{g,0} \ge 24$ mag arcsec$^2$, and effective radius, $r_e \ge 1.5$ kpc, probably returns a set of objects that has significant overlap with cluster dE's and dSph's that have been studied previously \citep[eg.][]{sandage,conselice,penny}. With time, providing the preliminary results on the few large UDGs already studied extend to a significantly sized population of similar objects, the UDG criteria may be refined to reflect a somewhat more distinct population that may not be either of low mass or exclusively in clusters. For a graphical overview comparing different galaxy populations, we refer the reader to Figure 12 of \cite{greco}.

We seek to identify as extensive a sample of physically large UDGs as possible across all environments to establish the characteristics of this population. The largest UDGs are extremely interesting for dark matter studies because: 
(1) they may be dark matter dominated at all radii;
(2) when compared to ultrafaint $(M_V > -8$ mag) galaxies, which are also dark matter dominated throughout, their expected large internal velocities ($> 30$ km sec$^{-1}$) can ultimately be measured to better relative precision; (3) unlike for the ultrafaint galaxies, there is a parallel sequence of galaxies (normal high surface brightness galaxies of the same total mass) that can be used to help us unravel the effects of baryonic physics on the dark matter distribution; (4) in contrast to dwarf galaxies, their large total mass makes their halos less susceptible to external tides and internal hydrodynamics; (5) they may have satellite systems of their own that can be used to study the dynamics at larger radii than can be probed with integrated light spectroscopy \citep[globular clusters have already been used in this manner, for example, by][]{beasleya}; and (6) they are found in all environments \citep[cf.][]{makarov,martinez,vdb17,roman17a,roman17b,greco,shi,wittman,leisman,prole}, enabling a comparison of dark matter halos across environment.

We have begun a survey that will increase the areal coverage and the expected number of large UDGs by orders of magnitude over what is available in the literature today. This work is possible due to the Legacy Surveys imaging data \citep{dey}, which are being obtained in support of the Dark Energy Spectroscopic Instrument project (Schlegel et al. 2011; DESI Collaboration et al. 2016a,b; http://desi.lbl.gov).
In this paper, we describe how we are automating the data analysis by presenting results from a relatively small portion of sky, a $\sim 10^\circ$ projected radius region around the Coma cluster, to describe our survey and demonstrate what might be expected of the full survey. We are aiming to ultimately make the intermediate steps public, in addition to the final catalog. We find that our processing generates UDG catalogs that are already competitive with the best published surveys in terms of identifying these large UDGs out to the distance of the Coma cluster. In \S2 we describe the publicly available data. In \S3 we describe our procedure for removing the visible, high surface brightness objects from the images, for smoothing the residuals to increase the signal-to-noise ratio (S/N) of diffuse sources, for identifying real sources, for distinguishing between UDGs and other astronomical sources that manifest as low surface brightness enhancements, and for obtaining parameter estimates. In \S4 we describe the findings, focusing on the distribution of UDGs both on the sky and along the line of sight within this volume, and the size and color distribution. We adopt a standard $\Lambda$CDM cosmology when necessary with H$_0 = 70$ km sec$^{-1}$ Mpc$^{-1}$, $\Omega_m = 0.3$, $\Omega_{\Lambda} = 0.7$.  To be consistent with \cite{vdk15a}, we assume an angular distance of 98 Mpc to the Coma cluster, which implies a physical to angular scale of 0.475 kpc arcsec$^{-1}$. The corresponding luminosity distance is 102.7 Mpc. 
All magnitudes are on the AB system \citep{oke1,oke2}.

\section{The Data}
\label{sec:data}

In preparation for the DESI program, the DESI collaboration has undertaken the Legacy Surveys, a deep 3-band ($g$ = 24.7, $r$ = 23.9, and $z$ = 23.0 AB mag, 5-sigma point-source
limits) imaging survey using DECam at the CTIO 4-m (DECaLS), an upgraded MOSAIC camera at the KPNO 4-m (MzLS, Mayall $z$-band Legacy Survey), and the 90Prime camera at the Steward Observatory 2.3m telescope (BASS, Beijing-Arizona Sky Survey). This survey will be roughly 2 magnitudes deeper than SDSS and is fully described by \cite{dey}. 
The original planned footprint has 9000 deg$^2$ observed by DECaLS and
5000 deg$^2$ observed by MzLS and BASS.
DECaLS, MzLS, and BASS data are already publicly available at the National Optical Astronomy Observatory (NOAO) science archives\footnote{\href{http://archive.noao.edu}{archive.noao.edu}} and the Legacy Surveys website (legacysurvey.org). Here, we present the 275 large (effective radii $r_e \gtrsim$ 5.3 arcsec, corresponding to physical values $\ge$ 2.5 kpc at the distance of Coma) UDG candidates identified within an $\approx$  10$^\circ$ radius of the Coma galaxy cluster using publicly released, reduced data from the DECaLS survey that were available as of April 1, 2018. In the future, our work will also include analysis of public BASS and MzLS data, but in this introductory study we focus exclusively on DECaLS data. 

DECam consists of 62 imaging CCDs arranged in a hexagonal pattern covering ~3 deg$^2$ with a pixel scale of 0.263 arcsec \citep{flaugher}.  During the period covered by this study 60 or 61 of the 62 CCDs were available for analysis.  Exposure times range from 40 to 250 sec, varying with observation conditions to reach the specific 5$\sigma$ point-source criteria in two out of three exposures.
We use the images made available online via NOAO, which have previously undergone calibration and processing (InstCal) via the DECam Community Pipeline \citep{valdes}. We utilize both the object and data quality masks for foreground subtraction and UDG detection as described further below.  Parameter values used in calculations such as exposure times, zero points, and pixel scales are obtained from the FITS headers.

\begin{figure*}[t]
\includegraphics[width=1.0\textwidth]{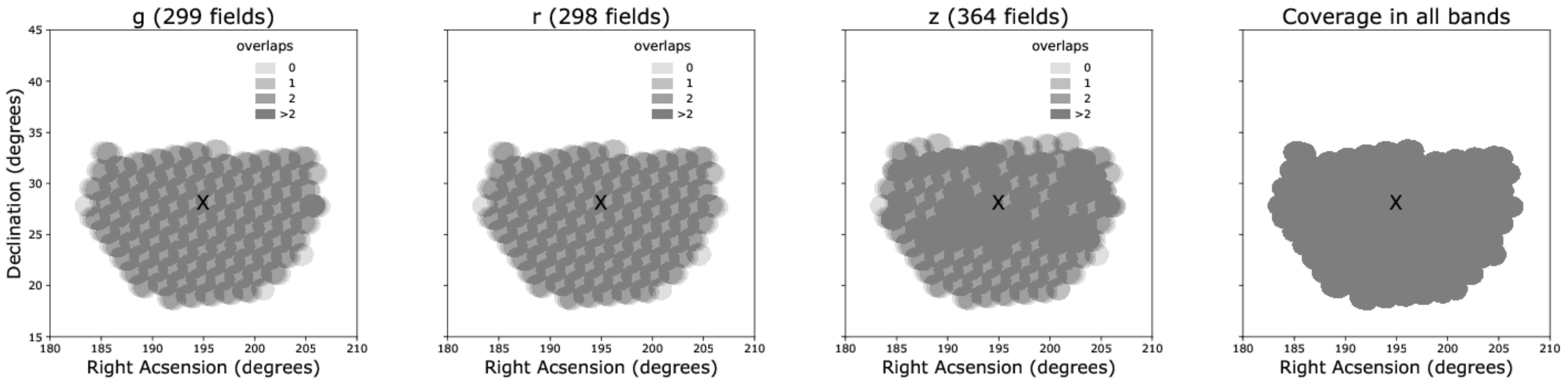}
\caption{The available data, as of April 1, 2018, in the $g$, $r$, and $z$ bands within 10$^\circ$ of the Coma galaxy cluster. In the first three panels, the greyscale denotes the number of the combined set of exposures. The rightmost panel shows regions for which we have coverage in all three filters.  The dark, central cross marks the center of the Coma cluster. }
\label{fig:footprint}
\end{figure*}

The Coma galaxy cluster lies near the northern boundary of the DECaLS survey footprint at a declination of about $+$30$^\circ$ and, as such, the 10$^\circ$ region around Coma that we target for study is cropped (Figure \ref{fig:footprint}). Nevertheless, the available data even within this truncated region, which is not yet the complete set intended for DECaLS, already consists of 974 separate DECam fields in three different filters ($g$, $r$, and $z$), of which 13 were flagged as bad exposures by the DESI collaboration. The remaining 961 fields (58320 separate CCD exposures) provide coverage for about 290 deg$^2$ in all three bands. We began with an analysis of the Coma cluster to enable direct comparison with previous UDG searches.

\section{Uncovering Diffuse Sources}

Various factors, some natural and some instrumental, limit our ability to locate low surface brightness galaxies. Previous studies have applied innovations to focus on different limiting aspects; examples include: drift scanning to address issues of image uniformity \citep{gonzalez}; specialized telescope baffling and new algorithmic approaches to address issues of scattered light \citep{mihos}; and refractive telescopes with new lens manufacturing techniques to mitigate the wings of the point spread function \citep{abraham}. Here, we face the principal challenge, given the use of archival data, that neither the 
instrumentation nor observations were optimized for the detection of low surface brightness galaxies. 

Despite this challenge, these data provide several key advantages over previous and ongoing surveys. First, such a large survey allows us to be relatively conservative in masking and still retain the ability to identify a large number of sources. Second, the multiple pass observational strategy provides some averaging over sky fluctuations. Third, the availability of images taken through three different filters, a luxury in low surface brightness surveys, enables us to confirm sources in different passbands and potentially to use color as a discriminant against contamination. Fourth, the relatively high spatial resolution ($\lesssim 1{\arcsec}$), which is typically not considered to be necessary when searching for large, diffuse sources \citep{vdk15a}, enables us to more effectively clean contaminating sources, particularly highly correlated, distant galaxies, which commonly produce low surface brightness enhancements in such searches \citep{gonzalez}. Finally, the Legacy Surveys cover most of the extragalactic sky, enabling the most comprehensive survey for UDGs from both hemispheres.

In this section, we outline our data analysis process and demonstrate that we are effectively complete in our detection of the galaxies found in the core of the Coma cluster using the Dragonfly telescope \citep{vdk15a} and, to a large degree, those found using the significantly more sensitive Subaru HyperSuprime images \citep{koda,yagi}. The majority of the galaxies that we 
miss from the \cite{koda} and \cite{yagi} studies are those that do not appear to match our selection criteria, not ones that we fail to detect. Therefore, our analysis of the DECaLS data will yield UDG candidates of the sizes and surface brightnesses highlighted in the \cite{vdk15a} study over thousands of square degrees of sky. 

The dominant source of {\sl effective} noise complicating the detection of low surface brightness objects in these images are resolved stars and galaxies. These can be exacerbated by resulting artifacts such as diffraction spikes and bleeds. In some cases the nature of source itself, such as the tidal tails and shells that surround some galaxies, complicates the identification of low surface brightness sources. Although aggressive masking reduces contamination from these foreground and background sources, such masking can also make it difficult to recover low surface brightness galaxies extending over relatively large angular scales.  Therefore, we take care to preserve low surface brightness, large-scale structures while reducing the effects of high surface brightness objects.  Our approach requires several steps to replace the extended wings of bright foreground and background objects with a model, rather than straightforward masking of large areas.  

\subsection{Image Preprocessing}
We carried out the image processing and analyses on the Ocelote
cluster at the University of Arizona High Performance Computing center \footnote{\href{https://docs.hpc.arizona.edu/display/UAHPC/Compute+Resources}{docs.hpc.arizona.edu/display/UAHPC/Compute+Resources}}.  Before removing all detectable foreground/background sources, it is necessary to reduce the dominant contaminating artifacts.  These include image defects, halos and ghosts of extremely bright stars, and amplifier differences.  Image defects (cosmic rays, bad pixels, bleed trails, etc.) already identified in the data quality mask provided by the DECam Community Pipeline are replaced with neighboring pixels that have been smoothed using a Gaussian filter, excluding the region of the defect. Noise is then added to these pixels using the statistical properties of the image background. 
Because each DECam CCD reads out using two channels \citep{ flaugher}, there are differences in gain and bias level between the two sides of the CCD. Although residual differences after pipeline processing are usually minor, even a small difference can make it difficult to characterize low surface brightness objects near the CCD center and become obvious when smoothing the data on large scales.  We compensate for this asymmetry by attributing  any measured difference to an error in the adopted bias level and add a correction to one side of the image.  We estimate the value of the correction by aggressively masking all objects on the image and computing the difference in the means in two 100 pixel wide strips along the CCD mid-line.  Extended wings of very bright stars can extend even beyond the masked regions and, therefore, this process is performed both before and after subtracting these wings and it is described further below.

We use the data quality mask and the process described next for detecting saturated stars and those with diffraction spikes.  First, we generate a problem star candidate list using SExtractor \citep[hereafter SE;][]{bertin} with per pixel detection threshold and minimum area settings of 2$\sigma$ and 500 pixels, respectively. We retain detections with a minor-to-major axis ratio $>0.8$. 
If the data quality mask flags saturated pixels within a candidate detection footprint, we classify it as a saturated star. A different approach is needed to identify stars without saturated pixels.  The four diffraction spikes generated by the telescope lie at about 45$^\circ$ from the vertical and horizontal CCD axes.  For each star candidate, we divide a region extending from 0.5 to 1.2 Kron radii into 16, 22.5$^\circ$ wedges starting at 11.25$^\circ$ from the long axis of the CCD.
Diffraction spikes will create a large variance among the mean fluxes within these wedges. 
If at least three out of four wedges containing potential diffraction spikes contain more flux than their two abutting neighbors, we classify the object as a problem star.  

Correcting for these problem stars is complicated because the profile of very bright objects is noticeably asymmetric and can be a function of the readout direction. Our process involves two steps. First, we measure the object centroid, axis ratio, and peak amplitude by fitting a Gaussian to its radial brightness profile between one and two Kron radii independently along the horizontal and vertical axes. We exclude pixels that are flagged in the data quality mask. Because we are only interested in defining centers and eccentricities at this stage, the model choice is not critical. Second, we use these results to fit the extended wings between 5 and 30 FWHM with an elliptical Moffat profile, which is then subtracted from the image. We present an example result in Figure \ref{fig:star}.

\subsection{Object Identification, Masking, and Removal}

\begin{figure*}[t]
\begin{center}
\includegraphics[width=0.9\textwidth]{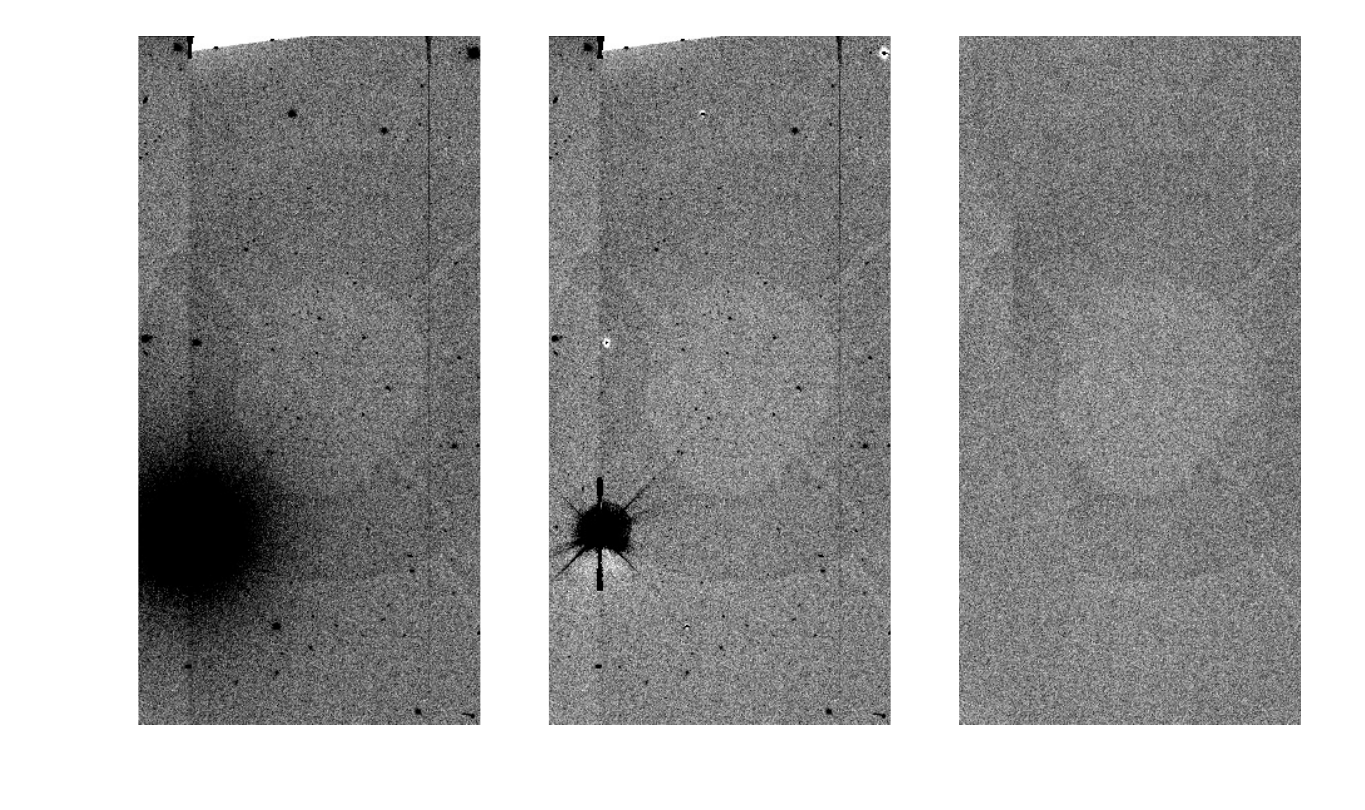}
\end{center}
\caption{The results after processing an image with an unusually bright object. The images span 18.3 $\times$ 9.1 arcmin.  Left panel:  The original CCD image containing a bright star with extended wings and associated pupil ghost.  Middle panel:  The extended wings of that star have been subtracted even for this bright an object. Right panel: The fully processed image shows the residual pupil ghost as well as faint evidence of the original star. The ghosts are difficult to model because they are field position dependent. Despite our processing, most of this image is inadequate for detecting faint UDGs, but fortunately such cases appear rare given our extensive visual inspection of candidates.  The bright feature in the upper left edge of the prior two panels was identified and replaced during the subtraction procedure.}
\label{fig:star}
\end{figure*}

Next, we subtract resolved foreground/background objects to remove sources of ``noise" identified in a smoothed image of the sky. While it is possible to aggressively mask these sources \citep{gonzalez}, 
such aggressive masking with the deep DECaLS images would eliminate too much sky and complicate UDG detection. At the other extreme, one could aim to model and subtract all detected sources. Unfortunately, it is impossible to model many bright sources well enough to produce accurate residuals at the flux levels of UDGs because galaxies tend to be irregular at such surface brightnesses. While we will never be able to identify a low surface brightness source coincident with a bright star or galaxy (Figure \ref{fig:star}), our goal is to subtract the wings of foreground/background sources sufficiently well that we can be more limited in our masking.

\begin{figure*}[t]
\begin{center}
\includegraphics[width=0.80\textwidth]{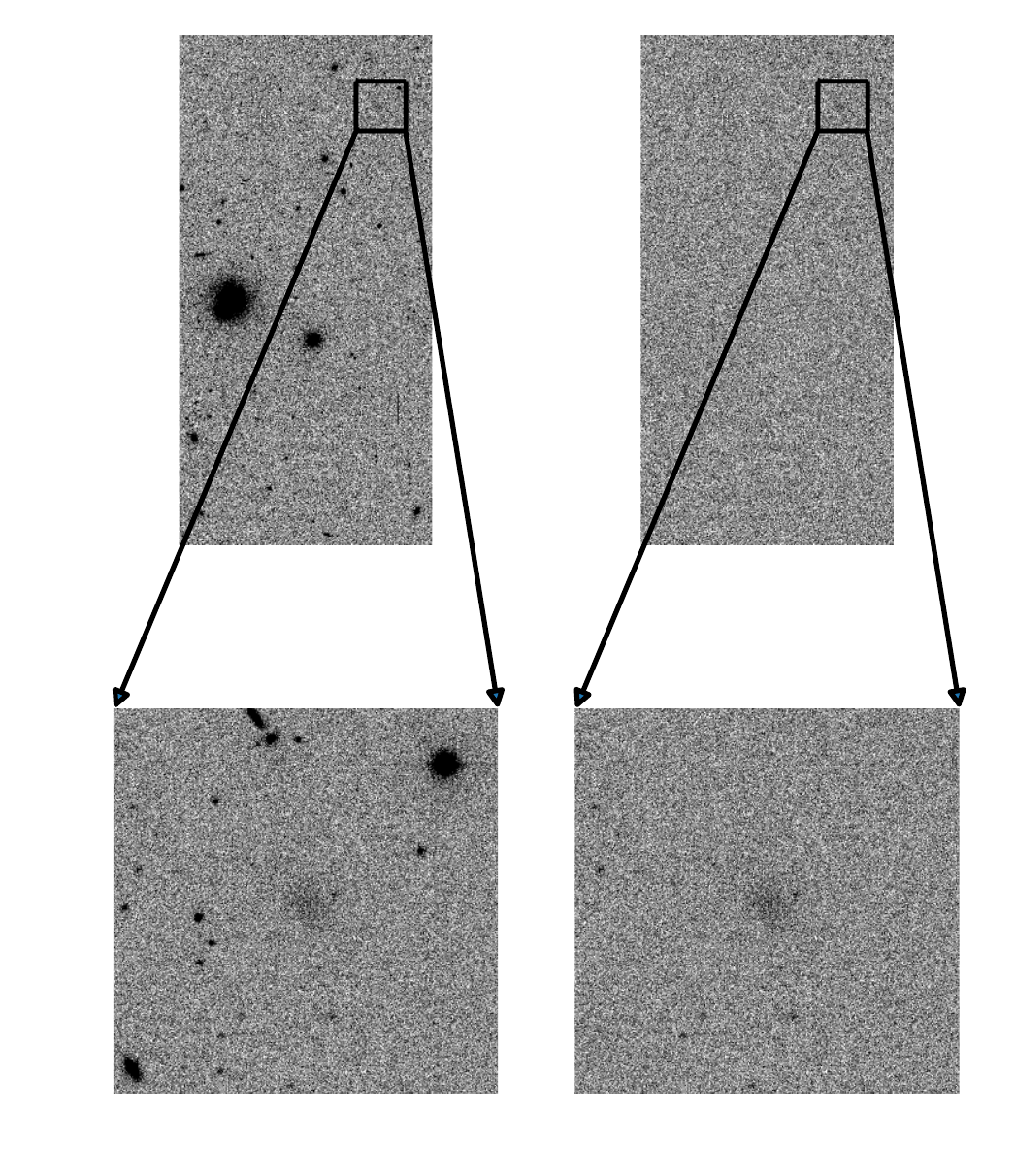}
\end{center}
\caption{An example of the entire subtraction process applied to one CCD image.  The upper panels (18.3 $\times$ 9.1 arcmin) show the image before and after high surface brightness objects have been removed.  The lower panels (108 $\times$ 108 arcsec) are exploded views of the regions outlined in the upper panels.  The linear stretch is the same in all panels.  The most noticeable residual object in the upper right panel is a UDG \citep[DF 07;][]{vdk15a} that lies at the center of the lower panels. Although the bright objects are removed from the lower right panel, several faint residual sources  were not identified and subtracted due to the chosen SE threshold levels. }
\label{fig:subtraction}
\end{figure*}

The objects that we do subtract are identified using SE with a per pixel detection threshold of 3$\sigma$ above sky and a minimum area of 6 pixels.  
These values are a compromise that enables detection of all but the faintest objects that are visible to the eye in the images and yet not detect UDGs. It is possible that these criteria, as well as others described below, will need modification when we examine fields with different source densities. All detections are initially replaced with a temporary elliptical mask with a size that is a function of its peak flux.  As discussed below, this temporary mask is eventually replaced with a model of the object's surrounding background.  We chose the mask size such that its edge has a value of 8 ADU above sky when fitted with an elliptical Moffat function. While this value produced good results over a range of S/N and different bands, it is not optimized for these parameters and will likely be adjusted after completeness studies. Its purpose is to limit the size of the mask while still allowing the source wings outside of the masked region to be approximated by an elliptical exponential function, $He^{-\gamma r}$, where $H$ is the peak flux and $\gamma$ is a scale factor.  The radius $r$ is calculated using
\begin{equation}
\begin{split}
r = [(\frac{\cos \theta^2}{a^2} + \frac{\sin \theta^2}{b^2}) (x-x_0)^2 + \\ (\frac{\sin \theta^2}{a^2} + \frac{\cos \theta^2}{b^2})(y-y_0)^2 + \\ 2\sin\theta\cos\theta(\frac{1}{a^2} -\frac{1}{b^2})(x-x_0)(y-y_0)]^{1/2},
\end{split}
\label{eq:radius}
\end{equation}
where $a$ and $b$ are the major and minor axes of the Moffat function, $\theta$ is its position angle, and $x_0$ and $y_0$ are the center coordinates.  The source is assumed to be superimposed on a tilted planar background, $S = Ax + By + C$, where $S$ is the sky level function, $A$ and $B$ are slopes along the vertical and horizontal axes, and $C$ is the mean sky level. Model fitting is done using the Python software package LMFIT \citep{newville}.  We subtract the source model from the surrounding region and replace the masked portion of the source with the tilted background derived from model plus noise estimated from the  image.  These approximations work very well for most galaxies and stars (Figure \ref{fig:subtraction}) although, as shown in Figure \ref{fig:star}, they can fail for unusually bright objects.  The average total replaced masked area for a CCD is 13.0\% but did reach levels of over 50\% in some regions such as the center of the cluster. While this process can leave  faint residual sources that are not UDGs in the image (Figure \ref{fig:subtraction}), the UDGs are unaffected. We refer to these processed images as the ``cleaned'' images.

\clearpage

\subsection{Wavelet Filtering}
\label{Wavelet Smoothing}

A key step in detecting UDGs is to spatially filter the cleaned images because UDGs, of the type detected so far, are not particularly faint in terms of total apparent magnitude. 
While bright objects have been removed from these images, low-surface-brightness, extended objects must still be detected among the many residual compact sources and background sky noise.  A general method for achieving this is to use smoothing to improve the overall S/N for extended sources by decreasing sky noise and then apply two-dimensional filters to enhance signals associated with objects of a particular size.
There are various approaches that can be taken here, including matched filtering if the parameters of the source population are fairly well constrained \citep[eg.][]{dalcanton,gonzalez}. Advantages of this approach are that sources with other characteristics are discriminated against and the smoothing is optimized for the intended sources. An alternate approach is ``thresholding", where one identifies pixels above a certain threshold level and identifies clusters of such pixels \citep[eg.][]{kniazev,bennet}. The advantages of this technique include that no prior knowledge (or bias) of the source population is necessary and that it is computationally efficient and simple to implement and model. We opt for yet a third approach.  

Wavelet transforms enable us to isolate objects at different scales with a tailored filter.  Their ability to spatially isolate objects by size and location in noisy fields has led to their application in astronomy for detection of faint objects in complex environments \citep{anisimova, damiani, starck}. We implement a multilevel, stationary wavelet transform using a version of the {\it{\`a trous}} algorithm as described by \citet{starck}. For our discrete smoothing kernel, $K$, we use the scaled difference of two-dimensional normalized exponentials
\begin{equation}
K = n\frac{e^{-r/k}}{2\pi k^2}- (n-1)\frac{e^{-r/2k}}{8\pi k^2},
\end{equation}
where $n$ is a scaling factor and $k$ defines the radial scale.
The first term can be thought of as the constructive term of the kernel that helps increase the S/N of a source of radial scale $r=k$ and the second term as the destructive term that diminishes the contribution from light at larger scales. This is reminiscent of the unsharp masking technique. The critical feature is that unlike smoothing this approach separates contributions from features at different scales. The scaling factor, $n$, determines both the steepness and the zero crossing of the kernel.  A value of $n = 1$ results in an exponential profile with no zero cutoff.  We choose a value of 1.5 which maintains a broad profile with a discrete zero crossing.  We use a radial scale, $k$, of 4.0 pixels for the starting kernel since it has a size profile (FWHM $\approx$  1.4$^{\prime\prime}$) on the order of the smallest objects detectable by DECaLS. New wavelet levels are created by upsampling the kernel in the previous level be a factor of two.  Therefore, the kernel for each level, $l$, is expanded by a factor of 2$^{l-1}$ (the baseline is designated as level one) from its baseline size. For a given scaling factor and radial scale, higher wavelet levels have broader profiles and will selectively detect larger objects.

\begin{figure*}
\includegraphics[width=1.0\textwidth]{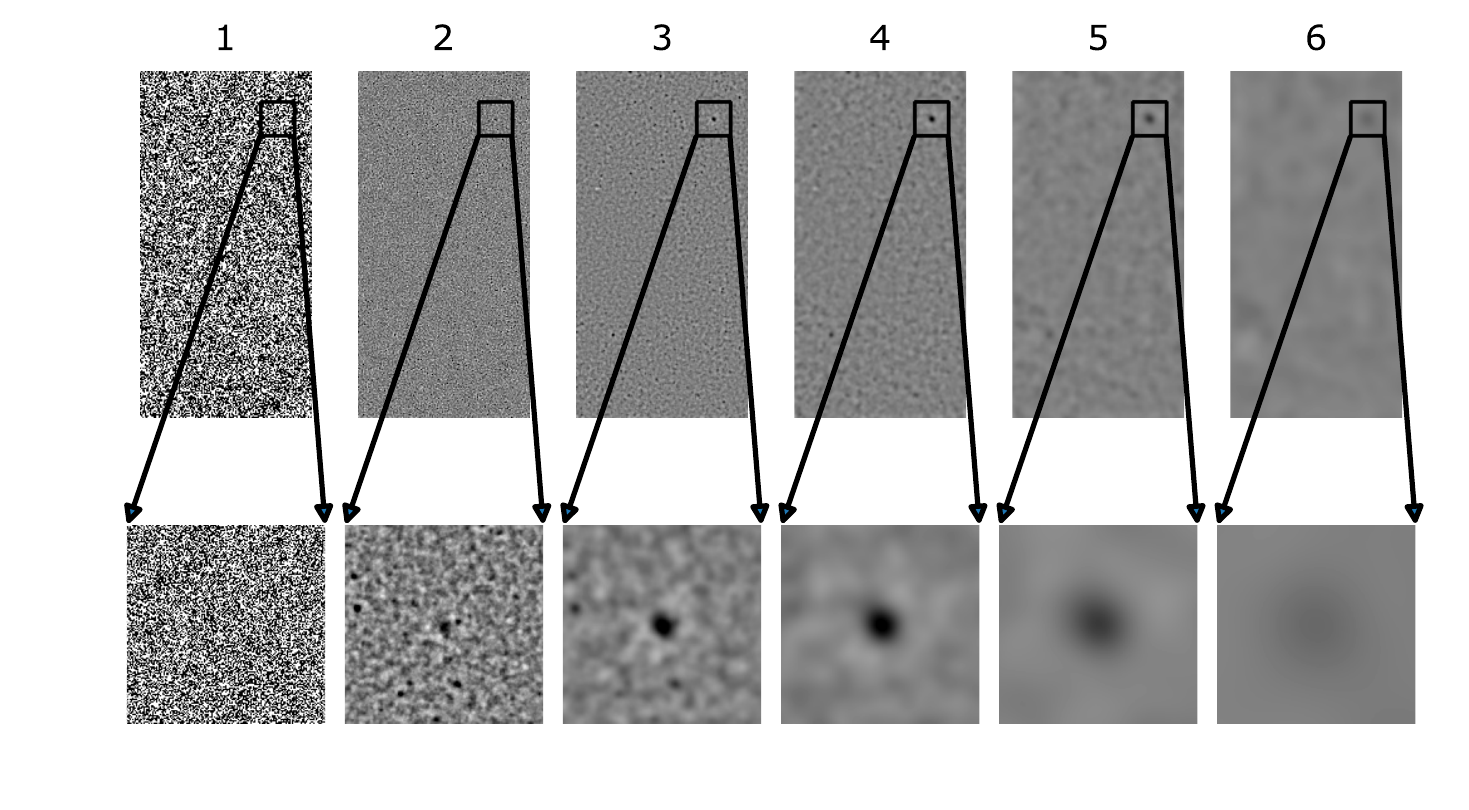}
\caption{An example of the wavelet process applying a range of levels  to the cleaned image from Figure \ref{fig:subtraction}. 
The numbers above each column represent the wavelet scaling level, $l$.  All panels are shown at the same intensity stretch. The contrast is manifestly maximized at certain scales, presumably those that better match the source angular size. Images in the upper panels span 18.3 $\times$ 9.1 arcmin and in the lower panels, 108 $\times$ 108 arcsec.}
\label{fig:waveletsEx}
\end{figure*}

An example of this process applied to the image from Figure \ref{fig:subtraction} is shown in Figure \ref{fig:waveletsEx}. The UDG is first visible at wavelet level two and becomes more prominent at larger scales, allowing it to be easily isolated from smaller objects.  We find that level four is appropriate for identifying UDG candidates of the size reported in this paper.  As seen in Figure \ref{fig:waveletsEx}, objects can be present at different wavelet levels. Relatively compact objects are more prominent in the images filtered at lower wavelet levels, whereas the more extended objects dominate at higher levels.  We take advantage of this filtering by using levels two and four to provide an initial list of potential UDGs presented here.

\subsection{Detection, Confirmation, Parametric Fits, and Object Classification}

\begin{deluxetable}{lrr}
\tablecaption{Screening Survivors}
\tablewidth{0pt}
\tablehead{
\colhead{Process}&
\colhead{Candidate galaxies}\\
}
\startdata
Wavelet screening & 3,574,596 \\
Object matching & 618,028\\
S\'ersic screening & 55,059\\
Galfit screening & 1079\\
Visual screening & 275\\
\enddata
\end{deluxetable}

Wavelet filtering and source detection is applied separately to each cleaned image.
To detect objects in the wavelet images, we first examine
the level four wavelet image.
A detection map is created from this image using SEP \citep{sep}, a set of libraries of stand-alone functions and classes based on SE \citep{bertin} that can perform many, but not all, of the functions efficiently in memory rather than using slower disk access.  This application, rather than SE, is utilized when we are primarily interested in only the detection image and background statistics.  We use a threshold of 2.5 times the background noise and a required minimum area of 300 pixels$^2$. These choices reflect a compromise between detecting the faintest objects and minimizing false detections. To  filter out small objects with extensions or merged neighbors, we create a separate detection map using the level two wavelet and a SEP threshold of 3 times background noise and a minimum area of 6 pixels. We require that any object present in both detection maps have at least 25\% of their peak flux in the level four detection map to be considered a UDG candidate.  We further require that the object be no closer than 30 pixels from the image edge, be unrelated to any saturated or bleed pixels flagged in the data quality mask, and have no more than 30\% of its area encroached by masks used in the initial object subtraction process.  This process results in an initial list of
3,574,596 UDG candidates (see Table 1) for
how various processing steps winnows down these candidates.

Candidates need to be confirmed as statistically real and then the objects need to be classified from among the full range of potential sources that appear as low surface brightness features in the images. 
We retain UDG candidates only if there are coincident detections among different exposures, regardless of filter, within 2\arcsec\ of the centroid of the group formed by the coincident detections. Each candidate is tested to see whether it is within 2\arcsec\ of an already existing grouping of detections or another isolated candidate. If it is related to an existing group, it is merged with that group and any outliers ($>$2\arcsec\ from the new centroid) are removed. Otherwise, it is merged with the paired detection to form a new group. To minimize duplicates, we compare UDG candidates within 4\arcsec\ of each other and retain only the one with the most individual detections. We will therefore be biased against separating very close pairs of UDGs if they exist. We found 618,028 groups (1,438,367 separate detections) fulfilling these criteria.

As defined by our science aims, our UDG candidates must meet specific size and surface brightness criteria.  To obtain rough but reliable estimates of these characteristics, which we will use to do a preliminary candidate screening, we apply the following procedure. 
We return to the uncleaned images to ensure that our processing did not remove a key component of the candidate UDG that could affect the measurement of its global characteristics. 
Our first challenge is that in many cases  
the UDG candidate, particularly if large, is not identified as a single object by SEP (i.e., having a single segmentation map identification number).  This problem is markedly reduced by generating a segmentation map using a smoothed version of the uncleaned image. The smoothing kernel we adopt is a Gaussian with a standard deviation of five pixels.  
Our second challenge is that there are often contaminating sources nearby. To mask these objects in the fitting, we will use the segmentation map obtained using the smoothed image, with one modification. We remove the object in the center of the segmentation map, presumably the candidate, from the mask.
Each detection is separately processed using a 201 $\times$ 201 pixel thumbnail image with the candidate in the center.  The segmentation map of the smoothed image is created using SEP with a detection threshold of twice the background noise and a minimum area of 10 pixels. These values represent a compromise between being able to detect the candidate and contaminating objects while avoiding excessive spurious detections. The central candidate is then identified and removed from the mask.  Using the \texttt{leastsq} function from the Python SciPy library \citep{jones}, an $n=1$ (exponential) S\'ersic model is fit to the unmasked pixels in the original uncleaned thumbnail. We choose $n=1$ to match the choice of \cite{vdk15a} for comparison and because that is broadly consistent with empirical determinations for similar objects \citep{conselice,yagi,vdb16}. The sky level during fitting is fixed as the mean background determined by SEP.

The resulting parameter estimates from our S\'ersic fitting are preliminary.  Although they are similar to those produced by GALFIT \citep{peng} later on, they are generated using only a single image detection rather than coadded images. To avoid rejecting detections that might meet our criteria when more precise estimates are obtained from the coadded images, we now set thresholds that are less strict than those that we will eventually require for our catalog.  To be retained as a UDG candidate at this stage, at least two individual detections must result in a fit with an effective radius $>$4\arcsec\ ($\sim$2 kpc at the distance of Coma) and a central surface brightness ($\mu_0$) greater than a value that depends upon the band.  Because we define a UDG as having a $g$-band central surface brightness $\ge$ 24 mag arcsec$^{-2}$, we set the current screening threshold in this band at 23.5 mag arcsec$^{-2}$.  Almost all UDGs will be brighter in the $r$ and $z$ bands and, to
avoid excluding potential UDGs in these bands, we set their screening thresholds at 22.5 and 22.0 mag arcsec$^{-2}$.  A total of 55,059 possible UDGs fulfill all of these criteria.

\begin{figure*}
\includegraphics[width=1.0\textwidth]{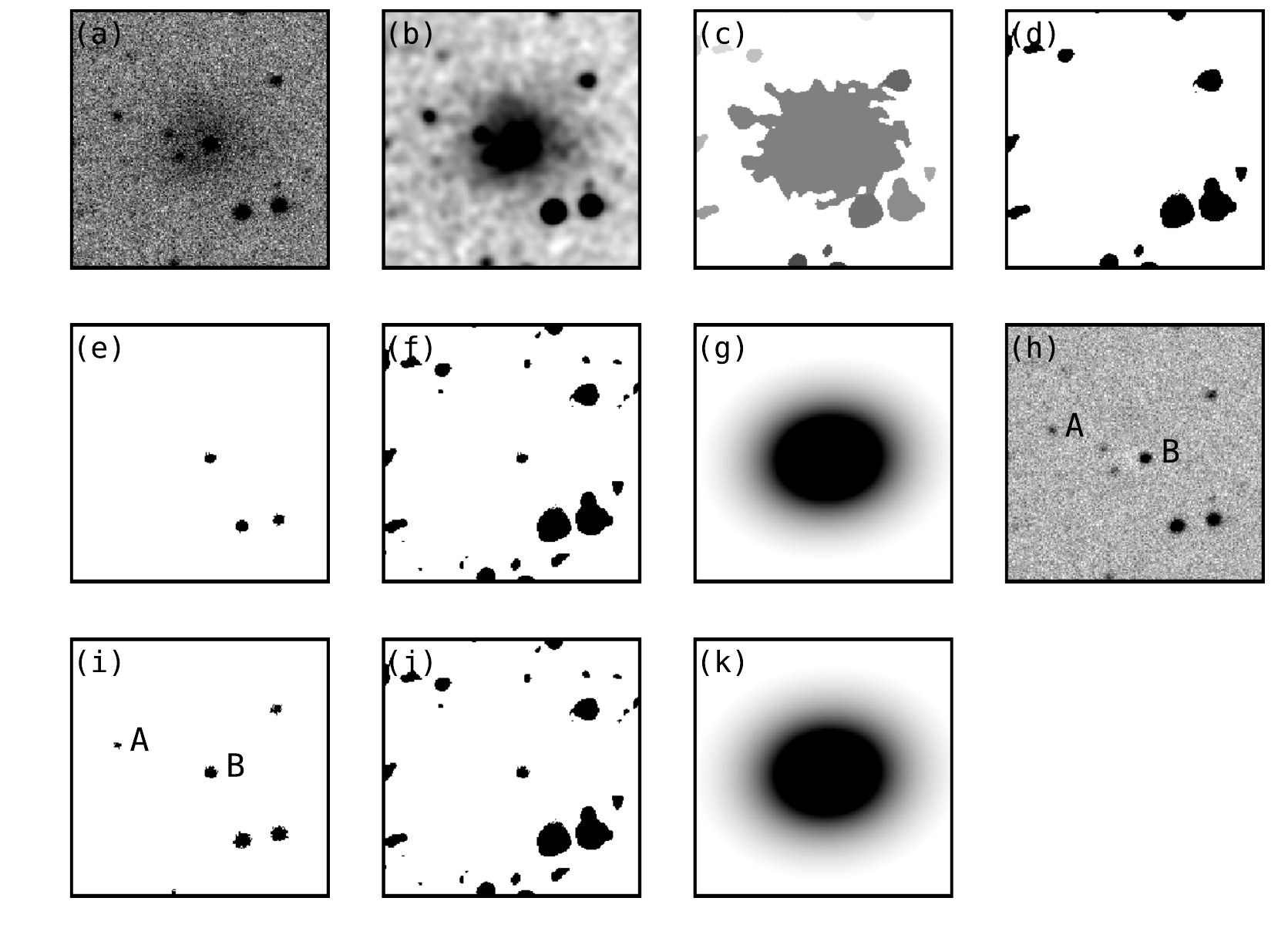}
\caption{Example of the steps involved in producing the GALFIT model.  All panels are 201 $\times$ 201 pixels (54 $\times$ 54 arcsec).  (a) Postage stamp of uncleaned image with the candidate in the center.  (b) A Gaussian smoothed version of Panel (a).  (c)  Segmentation map of smoothed image.  In this panel only, greyscale corresponds to the segmentation numbers.  In this case, the candidate is represented by the central, large detection.  (d)  The detection corresponding to the candidate is removed from the detection map (preliminary GALFIT mask).   (e)  Segmentation map of uncleaned, unsmoothed image required to detect overlying high surface brightness objects.   (f) A mask created by the union of Panels (d) and (f) which is used to create a  S\'ersic model of the uncleaned image.  (g) S\'ersic model  (h)  The residual after subtracting the  S\'ersic model from the uncleaned image.  (i)  SE segmentation map of the residual image.  (j)  A mask created by the union of Panels (d) and (i) which is used to create a  GALFIT model of the uncleaned image.  Both objects A and B shown in Panel (h) are within the candidate detection footprint.  In Panel (i) the SE FWHM of object A is $>$ 1.5 times the PSF FWHM, so it is removed from this final mask.  For object B, it is $<$ 1.5 times the PSF FWHM, so it is kept.  See text for details.  (k) Final GALFIT model.}
\label{fig:GalfitEx}
\end{figure*}

Next, we do a more detailed and precise examination of the surviving candidates using GALFIT, which is more versatile and provides more information than the basic S\'ersic fitting described above. An example of this process is shown in Figure \ref{fig:GalfitEx}.  We again return to the uncleaned images, as shown in Panel (a), to avoid a situation where part of the candidate was removed by our cleaning procedure. In this case, we simultaneously process all of the images containing the candidate, including those in which nothing was previously detected. We create coadded 201 $\times$ 201 pixel (54 $\times$ 54 arcsec) thumbnail images centered on the candidate for the analysis. Coadded images are both modeled separately for each filter and  using all images (full-stack). Once again we need to mask contaminating objects without rejecting faint structures within the candidate. Our first step in this process is the same as in our S\'ersic fitting process described above. We use a Gaussian filter to create a smoothed version of the uncleaned thumbnail and identify the candidate in the segmentation map (Panels (b) and (c)).   After separately masking negative outliers ($< -2\sigma$), we create a mask using pixels in the segmentation map that are not associated with the candidate (Panel (d)).   We refer to this as the preliminary GALFIT mask. 

The Gaussian smoothing helps keep the candidate intact during the segmentation process, but it also prevents us from identifying overlying high surface brightness objects. Therefore, in a second step, we create a mask of high surface brightness objects, including those superposed on the candidate (Panel (e)). We do this by creating a separate segmentation map of the uncleaned, unsmoothed thumbnail using a detection threshold of 1.5 times the central surface brightness threshold for a band (24.0, 23.6, and 23.0 mag arcsec$^{-2}$ for $g$, $r$, and $z$, respectively) with a minimum area of 10 pixels$^2$. This threshold is high enough to avoid detecting portions of a candidate that meet our central surface brightness criteria.  We combine this mask and that produced in the first step. This combined mask (Panel (f)) may include objects that are physically part of the candidate. Therefore, in a third step, we fit a S\'ersic exponential model (Panel (g)) to the unmasked regions of the cutout image (presumably containing only the candidate and remaining sky) and subtract the model from a copy of the uncleaned thumbnail to leave only the remaining overlying and peripheral objects (Panel (h)). 
Without additional information it is not possible to unequivocally discriminate between those objects that are part of the candidate and those that are contaminants. However, point sources, with the exception of a nuclear source, are more likely to be independent of the general structure of the candidate, even if physically associated, and so, we do not want to consider them during fitting.  Masking resolved sources superposed on the candidate could lead to masking of large areas therein, so we opt not to do so.

To implement this discrimination and create a new mask (Panel (i)), we use SE with a relative detection threshold of 2.5 and a minimum area of 10 pixels$^2$ to generate segmentation maps (for the different filter coadds and the full-stack) and measure the FWHM of each object in the residual image described above. In the coadded stacks of images in a particular filter, we define the point spread function (PSF) FWHM to be the maximum of the values obtained from the FITS headers of all CCDs contributing to the coadded image. For the PSF FWHM of the full-stack, we adopt the average of the PSF FWHM from the individual filter stacks. We assume that any object within the candidate footprint with a SE FWHM $>$ 1.5 times the PSF FWHM is part of the UDG candidate and it is removed from the mask. This mask does not adequately cover the wings of bright peripheral objects, so we combine this mask with the preliminary GALFIT mask described above (result shown in Panel (j)).

The unsmoothed, uncleaned image and the mask (Panels (a) and (j)) are supplied to GALFIT.  For defining UDGs, we elect to be consistent with the methodology of \cite{vdk15a} and use a fixed S\'ersic index of $n=1$ for the model with central surface brightness, $\mu_0$, calculated from the effective radius, $r_e$, the ratio of the minor to major axes, $b/a$, and the apparent total model magnitude.  However, as noted below, more general fitting is provided for photometry estimates.

A trimmed list of UDG candidates is generated by selecting those objects that satisfy 1) $\mu_{0,g} \ge 24$ mag arcsec$^{-2}$ or alternatively $\mu_{0,z} \ge 23$ mag arcsec$^{-2}$ if an observation in the $g$-band is not available, 2) $r_e \ge 2.5$ kpc at the distance of Coma, which corresponds to $\gtrsim$ 5.3\arcsec, and 3) $b/a \ge 0.37$. Both $r_e$ and $b/a$ are the values obtained from the fully stacked thumbnail. We selected the $b/a$ threshold as a compromise between limiting detections of artifacts (many of which, such as diffraction spikes, are narrow) without sacrificing a significant number of actual UDGs \citep[also see Figure 4b in][]{koda}. Only 1079 candidates satisfy these criteria.

\begin{figure*}[ht]
\begin{center}
\includegraphics[width=0.7\textwidth]{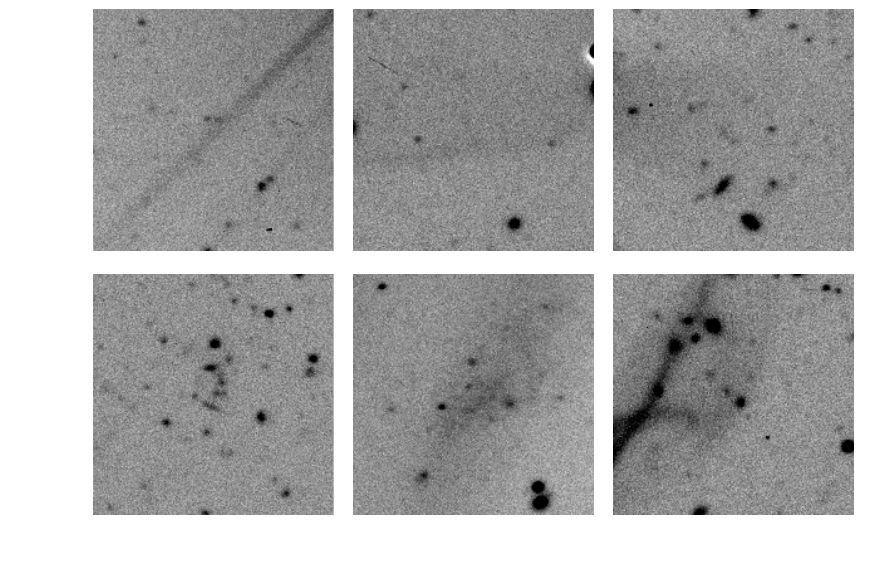}
\end{center}
\caption{Examples of detections rejected during initial screening.  Right ascension and declination of detections are shown in parentheses, following our visual classification.  Upper row from left to right: diffraction spike (184.4079, 28.9644), pupil ghost  (185.7454, 25.9153), and scattered light from nearby star (186.0171, 23.965). Lower row from left to right: cluster of galaxies (185.69, 27.5842), spiral arm of nearby galaxy (192.605, 25.4389), and tidal debris (190.6537, 25.7392).}
\label{fig:Rejections}
\end{figure*}

Our next step is to sort through the list for possible contaminants. Contaminants may result from artifacts, but may also be caused by astronomical sources such as extended stellar tails from galaxies, distant galaxy clusters, and nebulae within our own Galaxy. 
Ultimately, we need this screening to be automatic and reproducible. However, at first we need to develop our understanding of the resulting candidate sample and generate a training sample. To achieve both of these objectives, one of us (RD) visually classified all candidates, removing those that were confidently not UDGs from further review. The 650 rejected objects fall into three main categories 1) real objects such as peripheral parts of large, bright galaxies (tidal debris, spiral arms), clusters of stars or galaxies, and dust, 2) wings and scattered light from very bright galaxies and stars, and 3) instrument artifacts (pupil ghosts, diffraction spikes, CCD artifacts, etc.). 
Examples of rejected detections are shown in Figure \ref{fig:Rejections}.  Full images and surrounding sky may be seen using the Legacy Survey Sky
Viewer\footnote{\href{http://legacysurvey.org/viewer}{legacysurvey.org/viewer}}.  

\begin{figure*}[ht]
\begin{center}
\includegraphics[width=0.7\textwidth]{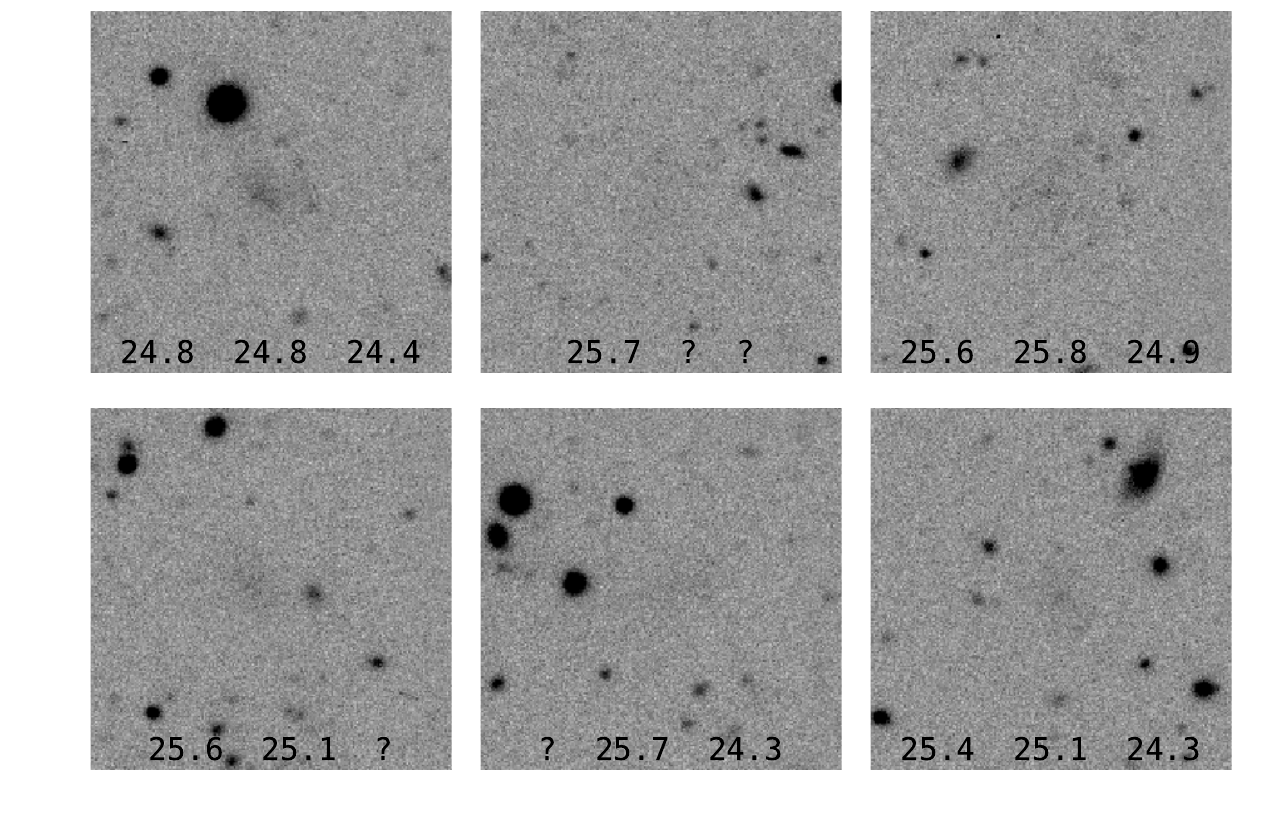}
\end{center}
\caption{Examples of candidates with conflicting reviewer classifications.  We present the strongest competing alternative classification to the UDG class with their names (if included in our catalog) and coordinates.  Numbers at the bottom of each panel represent $\mu_{0,g},\, \mu_{0,r}$, and $\mu_{0,z}$, respectively. A question mark indicates that no result is available for the associated band. All objects in the upper row were not ultimately classified as UDGs and are instead, from left to right, tidal debris (195.604, 26.8792), a background fluctuation (198.6044, 29.6972), and a faint galaxy cluster (192.9072, 27.345). All objects in the lower row were classified as UDGs, when we excluded that they might be, from left to right, tidal debris (SMDG1228096+300844; 187.0398, 30.1457), a background fluctuation (SMDG1311062+193524; 197.7758, 19.5901), or a galaxy cluster (SMDG1307184+184326; 196.8266, 18.7240). As is evident, at the limit of the data the classification is difficult.
}
\label{fig:Disagreements} 
\end{figure*}

Two of the authors (RD and DZ) examined the remaining 429 candidates in more detail. They disagreed on the classification of 63 (15\%) of these as an UDG, tidal debris, faint, high-redshift galaxy clusters, or  fluctuations in background sky (for very faint detections).  Arguments for accepting or rejecting objects from this group were independently reviewed by both examiners and any candidate not acceptable as a UDG to both was rejected.  Examples of candidates with initially conflicting classifications are shown in Figure \ref{fig:Disagreements}, where we also present  their final, consensus classification.  A total of 285 detections were classified as {\sl bona fide} UDG candidates. Our initial requirement that candidates be separated by at least 4\arcsec\ to be considered unique objects is inadequate for large UDGs and results in some duplication.  Therefore, we perform a final pass that requires that no object be within one $r_e$ of another candidate. This eliminates 8 duplicates, leaving us with a total of 277 potential UDGs.  

Because we want to automate classification using a machine learning classifier (see \S\ref{Automated Classification}), our visual decisions at this stage were based only on the available thumbnails without viewing surrounding fields, or other survey data, for additional clues. However, as demonstrated in \S\ref{Automated Classification}, this restriction can lead to erroneous assignments.  Because we do not want obviously incorrect classifications in our catalog, after finalizing the machine learning classifier, we re-examined all 277 potential UDGs in the Legacy Survey Sky Viewer. This review rejected two candidates, giving us 275 UDG candidates in our final catalog.  

While our initial GALFIT results are adequate for defining UDGs, more robust modeling is required for the type of photometric analysis presented in \S\ref{results}.  Therefore, new profiles are created of all 275 confirmed UDGs using GALFIT in a manner identical to above, except that the S\'ersic index, $n$, is allowed to float. 
The mean value of $n$ is 0.85, median is 0.73, and the standard deviation is 0.70.
 In some cases, our further analysis uses magnitudes and colors obtained from these models as specified. There is the danger of strong biases in the photometry of low surface brightness objects. 
These become particularly apparent at brightnesses significantly below sky level and can result in an overestimate of magnitudes while effective radii and S\'ersic indices are underestimated  \citep{haussler}.
 We will revisit the issue of optimizing the choice of photometric approach in upcoming work describing our recovery of simulated sources. For now, we only discuss  broad, qualitative first impressions from these data in \S4.

\subsection{Automated Classification}\label{Automated Classification}

Confirmation of UDG candidates has typically been done, both in this study and previous UDG studies \citep[cf.][]{vdk15a,koda,greco} on the basis of simple measurements or visual inspection. As documented in the previous section, this can be very time-consuming and subjective.  While visual classification is feasible with the limited area surrounding Coma discussed in this paper, it would be impractical for very large regions such as that included in the complete DESI footprint.  Moreover, because of the huge number of potential candidates, classification errors and subjectivity would make it difficult to replicate findings.  Finally, a fully automated classification strategy also allows for the processing of large numbers of artificial sources that can be used to assess completeness. Although we do not do that here, we do intend to insert artificial sources throughout our analysis of the full data set so as to have position-dependent completeness estimates. That work will be described in a subsequent paper. Of course, inherent subjectivity in the classification of the training sample will propagate to any automated technique. 

While machine learning methods have the distinct advantages of being able to classify images orders of magnitude faster than a human and provide consistent results, a potential major drawback is the reliability of their predictions.  Nonetheless, in some applications they have been shown to approach or even outperform human ``experts" in classification accuracy \citep{He}.  Modern algorithms initially developed for other purposes have recently been adapted for astronomical classifications \citep{ackermann, fowler, kim}. We briefly present our approach below.  A more detailed description of the methodology is  presented in Appendix \ref{app:classifier} and an extended treatment will be presented by Kadowaki et al. (in prep). Here, our main goal is to test the self-consistency of our visual classifications. Can an automated technique demonstrate that our visual classifications are at least internally consistent?

We implement our automated classifier with a modified version of the DenseNet-201 deep learning model \citep{huang} supplied in the Keras machine learning library\footnote{\href{https://github.com/keras-team/keras}{github.com/keras-team/keras}}. This model was designed to classify color images and requires three channels of data for each image and, therefore, we limit our dataset to those candidates with observations in all three filters (1071 of 1079 candidates). This is the full sample of candidates after GALFIT screening but prior to any visual classification. We set aside 20\% of the dataset (215 candidates with 160 non-UDGs and 55 UDGs) to be used for testing only after all network parameters are finalized. The remaining 856 images are used to train the network.  The accuracy after applying the trained model to the test set is 92.1\% with 9 false positives and 8 false negatives. These levels of accuracy are encouraging, but not necessarily optimized. In a subsequent study, we will explore alternative algorithms and compare results. For now, this degree of correspondence in our classification is adequate and suggests that our visual classification is internally self-consistent.

\begin{figure*}[ht]
\vskip 0.5cm
\begin{center}
\includegraphics[width=0.7\textwidth]{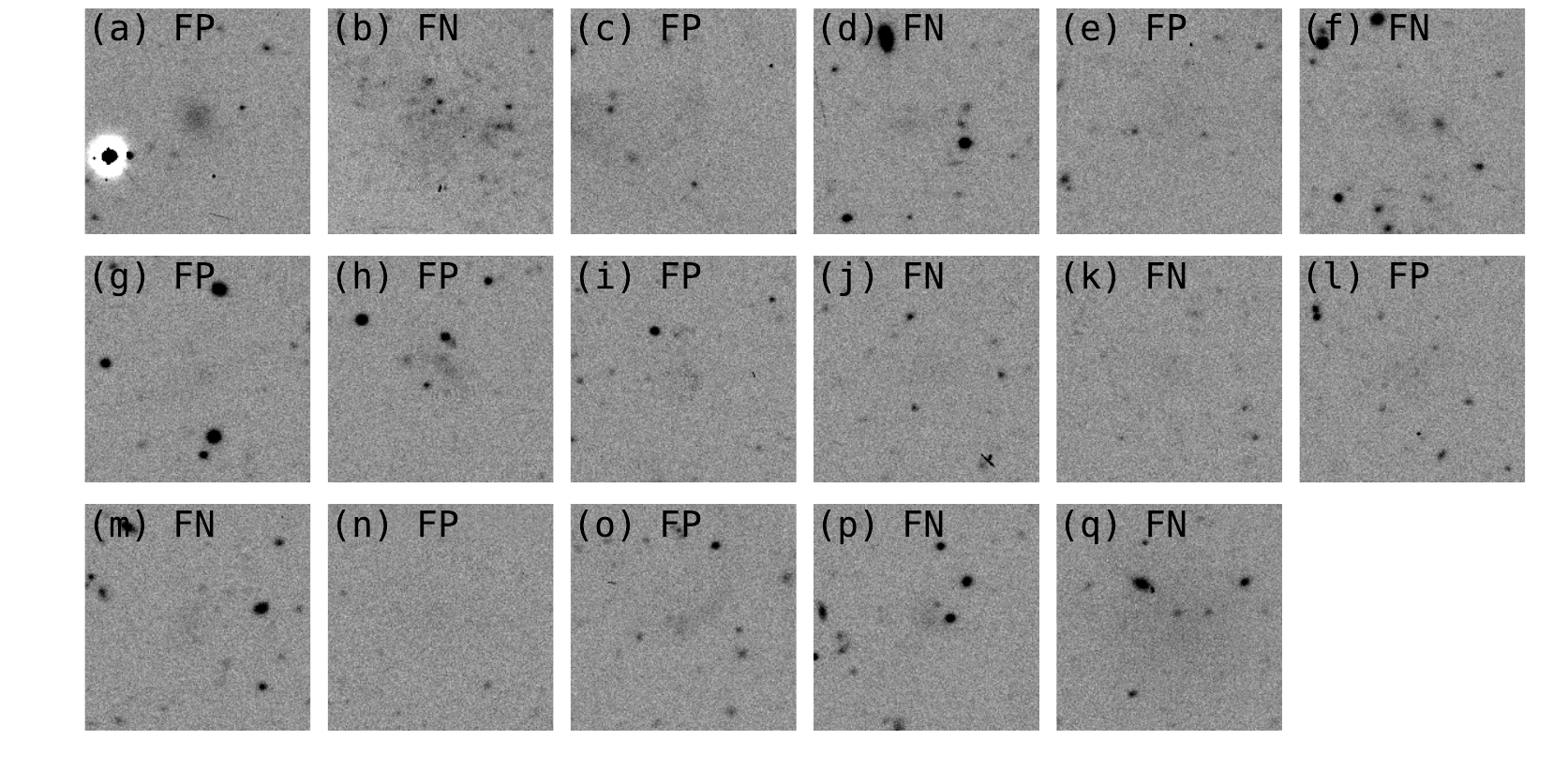}
\end{center}

\caption{Classifications where predictions produced by automated classifier differ from our visual classification. FN = False negative, FP = False positive relative to our visual classification. Right ascension, declination and names (if included in our catalog) are provided for each panel.  See text for details. (a) 194.7899, 29.0068 (b) 189.0129, 28.0035 (c) 184.1296, 30.8345 (d) 185.0787, 32.5963; SMDG1220189+323551 (e) 186.3063, 32.8416 (f) 187.04, 30.1449; SMDG1228096+300844 (g) 186.0624, 25.9412 (h) 186.3714, 24.2447 (i) 187.47025, 27.5274 (j) 194.2139, 27.2277; SMDG1256514+271338 (k) 195.5079, 28.0684; SMDG1302019+280407 (l) 196.3744, 23.4967 (m) 183.3408, 29.5664; SMDG1213219+293401 (n) 200.01857, 32.9523 (o) 202.7068, 27.3040 (p) 203.1018, 31.5587; SMDG1332244+313331 (q) 184.1619, 26.9951; SMDG1216389+265941}
\label{fig:Classifier} 
\end{figure*}

We show all of the candidates from the test set with predictions by the classifier that differ from our  visual classifications in Figure \ref{fig:Classifier}. Interestingly, we conclude that we mistakenly rejected the first candidate shown, (a), and that it was correctly identified as a UDG candidate by the automated classifier. In our initial inspection we classified suspected that it could be a distant galaxy cluster, but on re-inspection we agree with the automated classifier. Moreover, we visually classified another candidate, (b), as a UDG candidate based on reviewing only the thumbnails.  The classifier correctly rejected it and a larger image drawn from the Legacy Survey Sky Viewer clearly shows it to be an isolated segment of a spiral arm associated with a large galaxy.  We considered the faint object (c) to be an extension of a bright star just to the left of the thumbnail, although it appears to be separated from this region.  Panels (d) - (f) are among those that had conflicting classifications between the two human reviewers. These are generally  either very faint or could be composed of faint, high-z galaxies.

\subsection{The Catalog and Comparison with Previous Work}
\label{sec:comparison}

In Table \ref{tab:catalog} we present our final UDG catalog for this initial study of the region around the Coma cluster. This selection of UDGs is based on our visual classification, with minor adjustments motivated by the automated classification. In print we present an example of the full Table, showing only the first ten entries, while the associated electronic Table contains data for all 275 objects. We include coordinates both directly in the name of the object and in columns 2 and 3. In the next three columns we present the $g$, $r$, and $z$ central surface brightnesses. In the following two columns we present the structural parameters, effective radius ($r_e$) in arcsec, axis ratio ($b/a$), and major axis position angle ($\theta$) in degrees, where the latter is measured in the standard astronomical sense starting at North and measuring eastward on the sky. All of the aforementioned quantities are measured using a fixed $n$ Sers\'ic model, with the structural measures being measured from the full stack of available images. In the next three columns we present the apparent magnitudes measured using floating $n$ models. Uncertainties are those reported by GALFIT.  We will revisit these uncertainties using simulated objects in subsequent work. None of the magnitudes are extinction corrected.

One of the principal motivating factors for starting our work in the region of the Coma cluster is the existence of excellent prior surveys in the core of the cluster. These provide benchmarks against which we can test our false negative and positive rates, and our parameter estimation. We now describe our comparisons to two key samples.

\subsubsection{\cite{vdk15a}}
\label{vdk}

The catalog presented by \cite{vdk15a} based on the Dragonfly survey is dominated by the large UDGs that are also the focus of our survey.  
In Table 3 we present the cross-listed identifications for all of the UDGs presented by \cite{vdk15a} that have $r_e > 2.5$ kpc, for $r_e$ as presented by those authors. Of the 30 UDGs from that study that satisfy this criteria, our catalog includes all but four. Two of those four (DF 14 and 18) are excluded from our catalog because our fit to $r_e$ is smaller than 2.5 kpc (one of which, DF 14, just misses the cutoff with $r_e= 2.46$ kpc). During our visual definition of the training sample, DZ and RD disagreed on whether or not DF 35 is tidal debris. Because disagreements are treated as non-UDGs, this object is excluded from our final list (Table \ref{tab:catalog}). The final DF UDG that we excluded is DF 12. This is a more difficult case to resolve because we  rejected it on the basis of a low b/a, but on examination we find that the fitted model is poor and likely includes a nearby object. This is the one case of the four that is probably an incorrect exclusion on our part; however, because we cannot visually inspect all of the fits for the full SMUDGES survey, we exclude this object from our catalog and treat it as a false negative. We stress here that we did not miss any DF object of sufficient size (2.5 kpc) on the basis of a surface brightness limit on our part. We conclude that the Legacy Surveys data is suitable for UDG work at a competitive level with dedicated low surface brightness surveys (e.g. Dragonfly). 

\begin{figure*}[ht]
\vskip 0.5cm
\begin{center}
\includegraphics[width=0.7\textwidth]{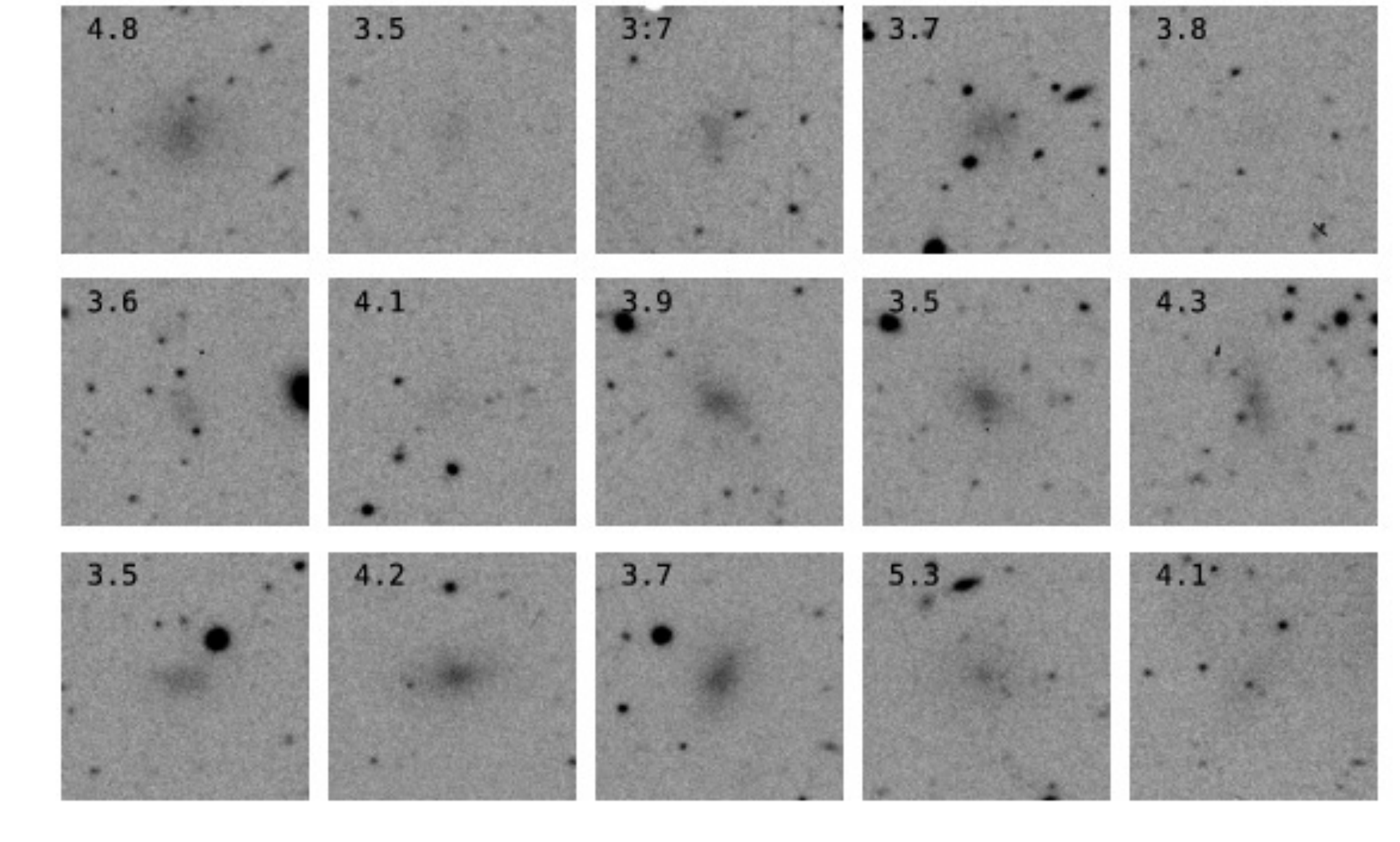}
\end{center}
\caption{Our detections that lie within the DF footprint and have $r_e > 3.5$ kpc that are not include in the DF catalog. The measured $r_e$ is given in the upper left corner of each panel. The UDGs (with coordinates) in order from top left to right are SMDG1253489+273934 (193.4536, 27.6594), SMDG1254345+274046 (193.6439, 27.6796), SMDG1255076+274407 (193.7816, 27.735), SMDG1256265+285928 (194.1106, 28.9915), SMDG1256514+271338 (194.2139, 27.2278), SMDG1257113+273405 (194.2966, 27.5681), SMDG1257446+280907 (194.4357, 28.1515), SMDG1258132+272522 (194.5548, 27.4229), SMDG1259413+273148 (194.9222, 27.5299), SMDG1300279+283730 (195.1164, 28.6251), SMDG1301124+274525 (195.3014, 27.757), SMDG1301158+271238 (195.316, 27.2107), SMDG1303312+285716 (195.8799, 28.9546), SMDG1303551+282120 (195.9795, 28.3555), and SMDG1305016+290859 (196.2562, 29.1496). Boxes are 54\arcsec\ ($\approx 26$ kpc at the distance of Coma) on a side.}
\label{fig:dfmisses}
\end{figure*}

It is also valuable to determine if we detect objects within the common survey footprint that \cite{vdk15a} did not. To be conservative in the sense that $r_e$ measurement errors would not  confuse the comparison, we consider only systems for which we measure $r_e > 3.5$ kpc.  With this requirement, we find
15 objects in our catalog that lie within the DF footprint but that are not in their catalog (Figure \ref{fig:dfmisses}). They noted confusion with the intracluster light as limiting their detections in the central regions of the cluster, but none of these 15 are within that region. They also emphasized isolated sources, so some of these may have been excluded as non-isolated. As shown in the Figure, these are mostly  unambiguous detections, so the reason for their absence in the DF catalog is unclear without a closer examination of their images, but in some cases they may have exceeded the central surface brightness criterion in the DF evaluation. For five of the fifteen we measure $24 < \mu_{0,g} < 24.5$ mag arcsec$^{-2}$, the remainder are of lower central surface brightness.

\subsubsection{\cite{yagi}}

This is the largest existing sample of UDGs in the Coma cluster, although the gain in size comes mostly from identifying smaller and fainter systems than those found by \cite{vdk15a}. For sources that do match our size criteria, we present 
the cross-listing in Table \ref{tab:koda}.
There are two principal ways in which objects in the Yagi catalog can fail to be in ours. First, we do not detect the source or it fails the basic criteria, such as satisfying the parameter requirements from the initial S\'ersic fit. Of the 75 objects in the published catalog that meet the size criteria, according to the effective radii presented by \cite{yagi}, 
15 are absent from our catalog and are identified as such in the caption of Table \ref{tab:koda}. Of those, two (numbers 199 and 456 in the \cite{yagi} catalog) are sufficiently bright (22.9 and 22.4 mag arcsec$^{-2}$, respectively)  that they were probably cleaned out of the images by our procedure. Four are sufficiently close to a bright object that they were either masked or removed as part of that object (231, 240, 257, and 370). Three were detected but failed the initial S\'ersic fit screening (4, 218, 471). Finally, five are sufficiently faint or possible extensions of a nearby source that it is difficult to confirm them in our data (94, 310, 425, 453, 569). 
None of these source has a corresponding SMDG number and so they are not included in the Table.

The second principal way in which objects in the Yagi catalog can fail to be in ours is if we rejected the candidate on the basis of our criteria. There are 26 such cases in Table \ref{tab:koda} and we describe the reason for their exclusion from our catalog. 
For example, 21 of these were, according to our measurements, either too small or have a central surface brightness that is too bright. We conclude that despite the deeper imaging available for the \cite{yagi} study, our survey for large UDGs is nearly as complete, with only a handful of examples of missed objects that could be attributed to our shallower image depth.

\subsection{Remaining Problems and Limitations}

None of the steps in our procedure is necessarily fully optimized, but certain aspects are perhaps easier to improve upon than others.
The procedure we describe consists of three basic phases. To recapitulate, first we remove as many sources from the images as possible to eliminate contamination for our smoothing or filtering of the data. Any large scale residual features in our final cleaned images arise principally from issues that would require adjustment of the telescope and camera optics or observing procedures. Therefore, we conclude that this first phase is reasonably close to being optimized and do not anticipate significant gains from alternate approaches as applied to these same data. Second, we use wavelet filters to increase the S/N of low surface brightness features. There are a number of alternate approaches and we suspect that there may be significant improvements possible in this step. Nevertheless, we are nearly complete in detecting known sources with the desired characteristics, so the incompleteness for these objects must be low. Improvements in this phase would perhaps allow us to reach even further into the low surface brightness population. Finally, the third phase is the classification of candidates. This step is important for sample purity and completeness. Our automated approach is an improvement over some coarse screening methods but will certainly improve once we have larger samples of confirmed, physically large, UDGs. As we improve our search techniques, there will be natural refinements in this last step. We anticipate that for our final catalog the classification criteria and efficiency will change from that presented here.  Completeness simulations will form an integral part of the final catalog.

A final complication on the completeness issue that must be acknowledged is that of clumpy, low surface brightness objects. Our artificial objects are smooth by construction and we know so little about the internal structure of UDGs that we cannot yet create models of clumpy UDGs. We expect to detect UDGs that might have modest star forming clumps, as long as they also have an underlying smooth stellar population, but estimating completeness for clump-dominated objects will be difficult. 

Within the survey regions described here, we have not faced strong contamination from IR cirrus. Deeper surveys for low surface brightness detections have already been plagued by such emission \citep[e.g.][]{duc} and we expect that some regions of the Legacy Surveys will also be too difficult for our work. We do not yet have an estimate for how large a fraction of the full survey may be affected.

\section{Results}
\label{results}

\begin{figure}[ht]
\begin{center}
\includegraphics[width=0.4\textwidth]{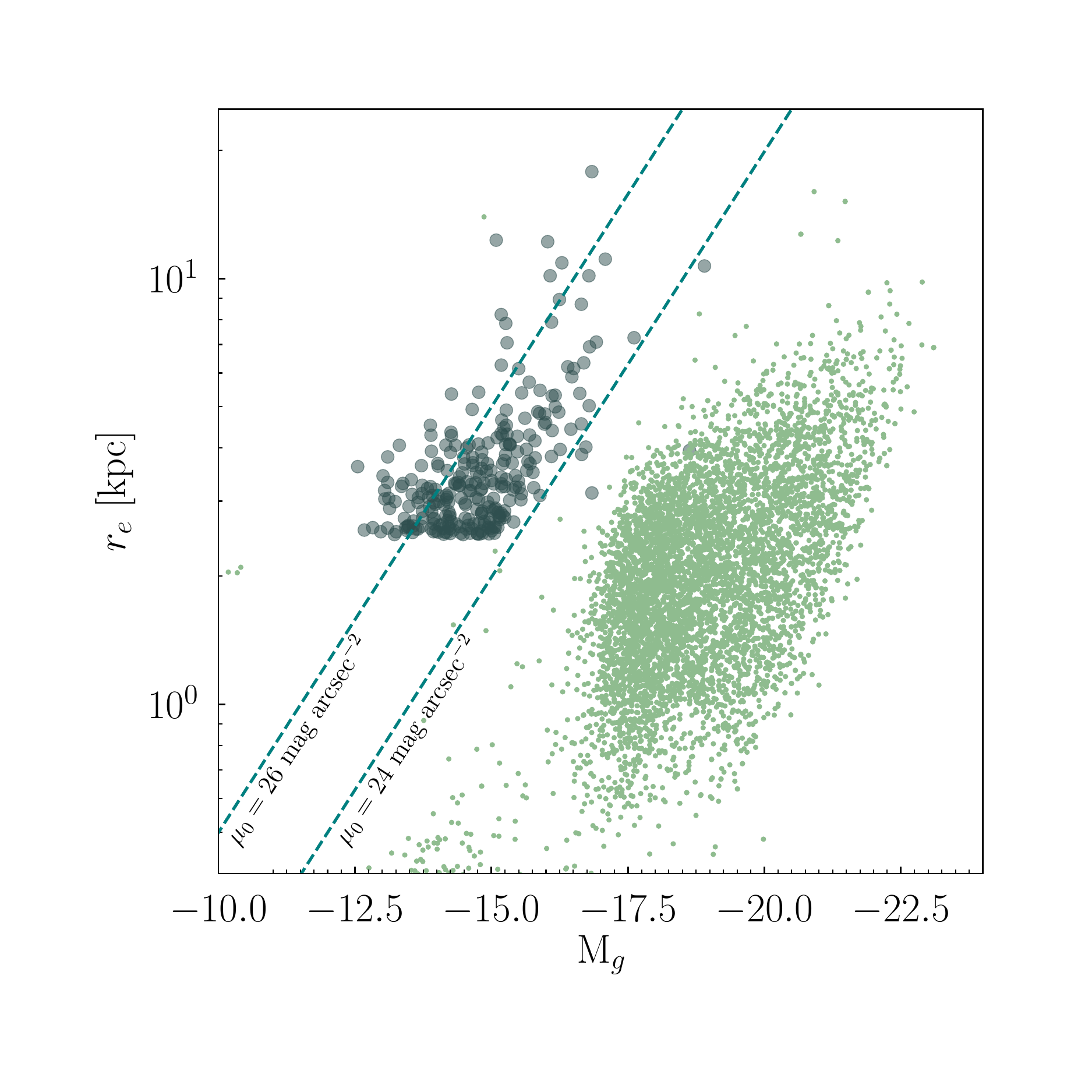}
\end{center}
\vskip -1cm
\caption{Absolute magnitude - size relation for the SDSS galaxies within the survey region and for $cz < 10000$ km sec$^{-1}$(smaller, lighter points) and the UDGs (larger, darker points). The absolute magnitudes of the UDGs are estimated by adopting a luminosity distance of 102.7 Mpc, consistent with Coma membership, for all. An incorrect distance assumption will slide a galaxy along the constant central surface brightness lines (dashed). The lines represent the loci of UDGs with b/a=1 and central surface brightnesses of either 24 or 26 mag arcsec$^{-2}$ in the g band.}
\label{fig:selection}
\end{figure}

The UDG catalog we present contains 275 UDG candidates that have  $r_e \gtrsim 5.3^{\prime\prime}$, which corresponds to $\ge$ 2.5 kpc at the distance of Coma, and central surface brightness $\mu_{0,g} \ge 24$ mag arcsec$^{-2}$ (Table \ref{tab:catalog}). The distribution of candidates in the absolute magnitude-size plane (assuming that they lie at the distance of the Coma cluster) is presented in Figure \ref{fig:selection}. By construction, there is little overlap with the SDSS sample of galaxies, showing that this is almost exclusively a previously unknown population of galaxies over large areas of the sky, with the exception, of course, of previous UDG work in this area of sky (\S\ref{sec:comparison}) and deeper surveys both in Coma and other clusters \citep[eg.][]{graham,vdb16} and the field \citep[eg.][]{greco}.
To place our UDGs in this figure, we assume that they lie at the distance of the Coma Cluster. Distance errors translate to motion of any individual galaxies parallel to the lines of constant surface brightness in the figure, and therefore do not affect the previous conclusion. On average,  $\mu_{0,g} \sim$ 25 mag arcsec$^{-2}$, with a tail of objects reaching to $\sim$ 26 mag arcsec$^{-2}$. Definitive conclusions regarding the size or magnitude distribution await a better understanding of the distances and the completeness across the parameter space. 

\subsection{Distance Constraints}

\begin{figure*}[ht]
\vskip -2cm
\begin{center}
\includegraphics[width=0.7\textwidth]{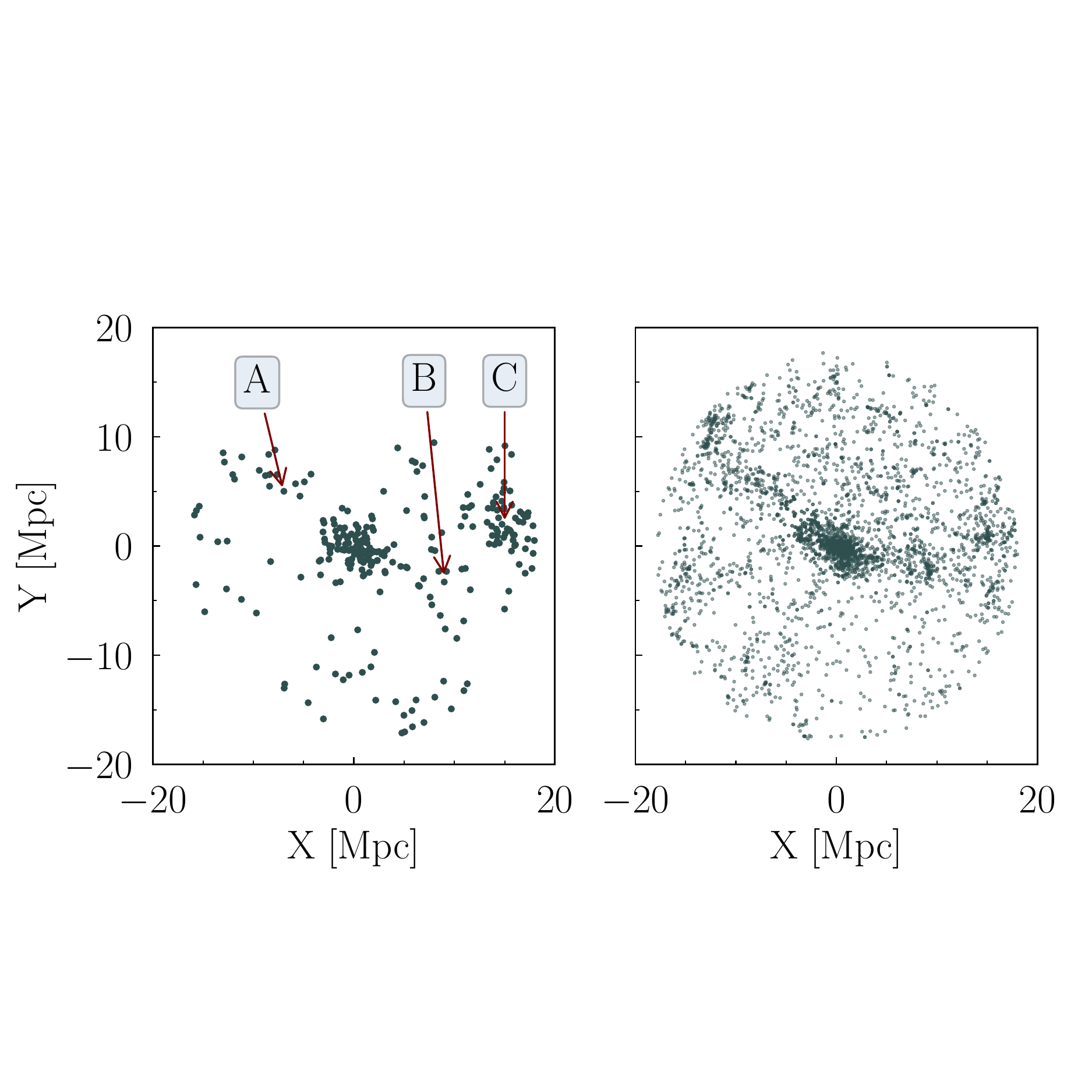}
\end{center}
\vskip -2cm
\caption{The projected distribution of UDGs (left panel) and SDSS galaxies (right panel) on the sky centered on the Coma galaxy cluster. We plot in physical units assuming all galaxies are at the Coma distance to provide guidance on the scale of the visible structures, which are at the distance of Coma. The UDGs represent our entire sample of 275. The SDSS galaxies include only those with recessional velocity $cz< 10,000$ km s$^{-1}$ and within a projected separation of 18 Mpc from the center of the Coma cluster. In addition to the clear clustering of both UDGs and SDSS galaxies in the Coma cluster, there are three other features in common that are labeled and discussed in the text. North is up and East to the left. On the sky, this figure represents a region with a radius of $\sim 10^{\circ}$. Our UDG survey has no data at Y $\gtrsim$ 10 Mpc (see Figure \ref{fig:footprint}). The lettered labels refer to structures discussed in \S4.1.}
\label{fig:posplot2}
\end{figure*}

\begin{figure}[ht]
\begin{center}
\includegraphics[width=0.4\textwidth]{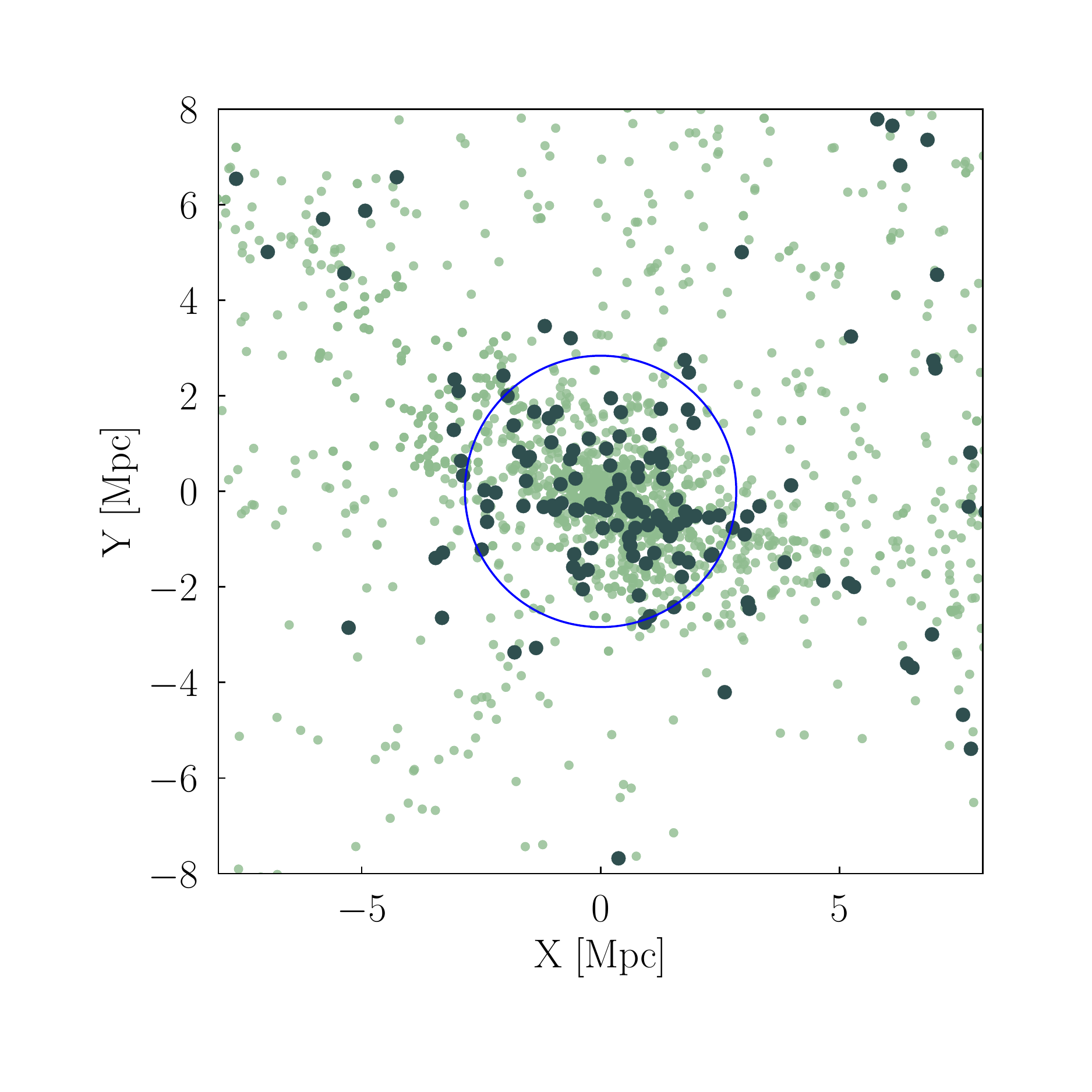}
\end{center}
\vskip -1cm
\caption{The projected distribution of UDGs (darker circles) and SDSS galaxies (lighter circles) on the sky centered on the Coma galaxy cluster. We plot in physical units assuming all UDGs are at the Coma distance. The SDSS galaxies include only those with recessional velocity, $cz$, $5500 < cz < 8500$ km s$^{-1}$. The blue circle represents the Virial radius, as reported by \cite{kubo}. North is up and East to the left.}
\label{fig:poscoma}
\end{figure}

Published spectroscopic redshifts, and therefore distance estimates, exist for fewer than 20 Coma UDGs and only a few outside of Coma \citep{vdk15b,kadowaki17,vdk17,alabi}. We continue to work to expand this sample both using optical spectroscopy (Kadowaki et al., in prep) and radio observations of the 21cm line to possibly detect the HI rich UDGs within this same area of sky (Karunakaran et al., in prep). However, this work will always be limited to  
a small fraction of all identified UDGs. 
We must address the challenge of understanding what the 2-D distribution of UDG candidates on the sky (Figure \ref{fig:posplot2}) implies for the true 3-D distribution and how best to estimate the distance to at least a subset of UDGs.

The relationship between the physical and apparent clustering properties of galaxies provides a constraint on the redshifts of a population \citep{rahman}. At its simplest, this argument underpins the interpretation that the bulk of the original Coma UDGs \citep{vdk15a} lie at the distance of the Coma cluster because they so evidently congregate around the galaxy cluster. This interpretation has been confirmed with a small number of spectroscopic redshifts \citep{vdk15b, kadowaki17, alabi}. We find a similar overabundance of UDGs in our survey (Figure \ref{fig:poscoma}), extending further from the Coma center to at least the virial radius, roughly 3 Mpc \citep{kubo}. In analogy to the previous claims, it is plausible to conclude that the bulk of these UDGs are also at the distance of Coma. 

\begin{figure}[ht]
\begin{center}
\includegraphics[width=0.5\textwidth]{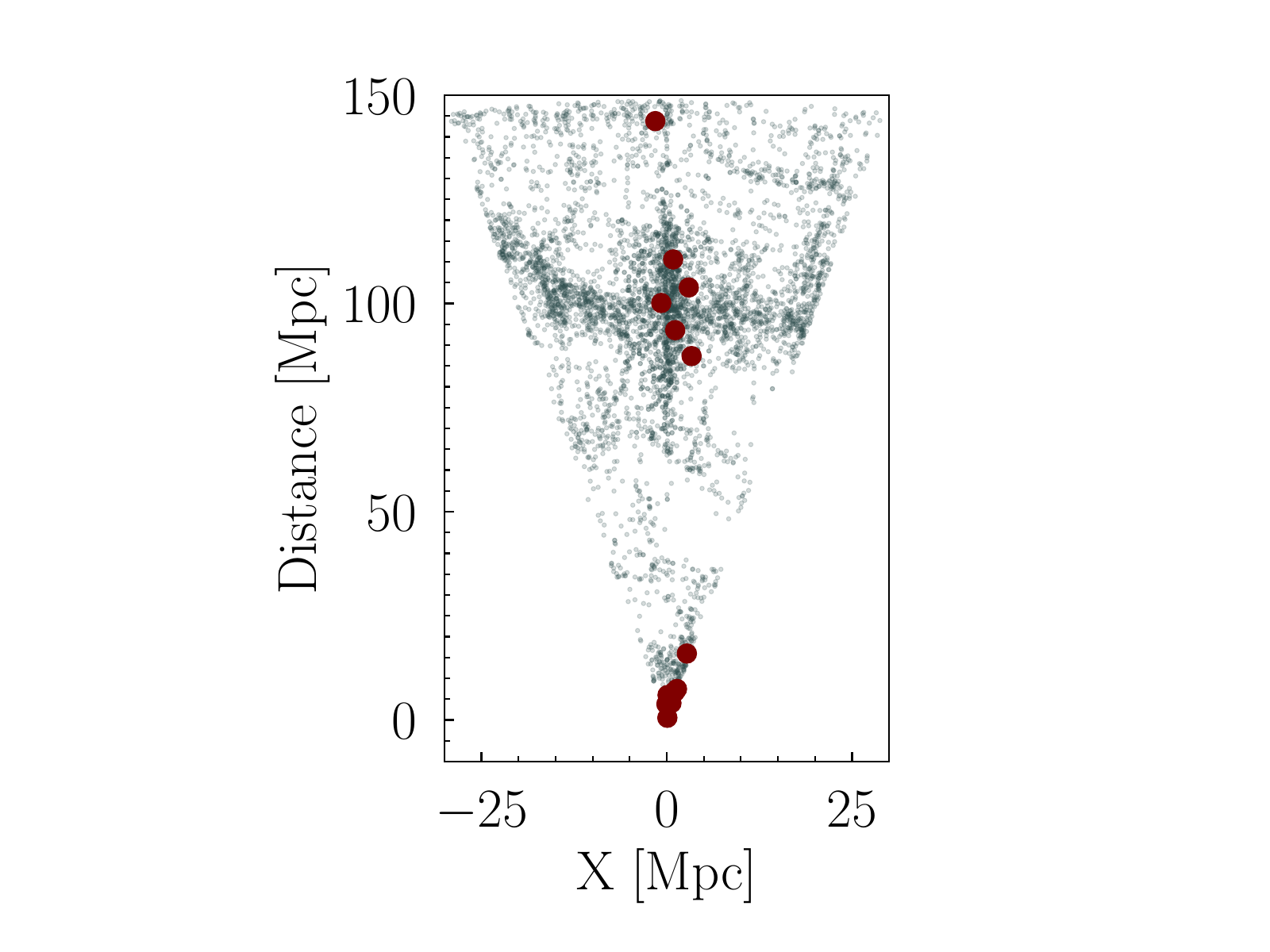}
\end{center}
\caption{The SDSS galaxy wedge diagram for the Coma region. The abscissa corresponds to projected distances from the Coma cluster along Right Ascension. The ordinate corresponds to Hubble flow velocity distances assuming H$_0 = 70$ km s$^{-1}$ Mpc$^{-1}$. The observer is at the lower vertex and the stretching of the Coma cluster along the line of sight is the well-known ``finger of God'' effect. East is to the left. Large circles show the distances of individual UDGs determined from spectroscopic redshifts, obtained either by \cite{kadowaki17}, Kadowaki et al. in prep, or the NASA Extragalactic Database (NED).}
\label{fig:zplot}
\end{figure}

Beyond the immediate vicinity of Coma, it is more difficult to associate UDGs with specific features seen in the normal galaxy distribution (Figure \ref{fig:posplot2}). There are locations where there is concurrence between the distribution of UDGs and SDSS galaxies, and other locations where there is not. We label three of 
the interesting features as A, B, and C in the Figure. Features A and C in the UDG distribution appear to match local large scale structure features near Coma.
These are analogous to the structures observed in the UDG distribution within the Abell 168 field \citep{roman17a}. They allow us to define subsamples beyond the Coma cluster for which distance-by-association is possible.
However, there are two concerns. First, feature B, which on the basis of the SDSS galaxy density one might have expected to be as evident as feature A, is not readily visible in the UDG distribution. Furthermore, feature C seems much richer in UDGs than a visual impression of the SDSS distribution might suggest. Such variations could point to interesting physical differences in the relative abundance of UDGs, or to errors introduced by associating UDGs too closely to their brighter cousins. In particular, for feature C, which is the result of the projection of sources along the right edge of the redshift wedge diagram (Figure \ref{fig:zplot}), contributions come from a local overdensity, a broad overdensity at the distance of Coma, and an overdensity somewhat beyond Coma.

\begin{figure}[ht]
\begin{center}
\includegraphics[width=0.5\textwidth]{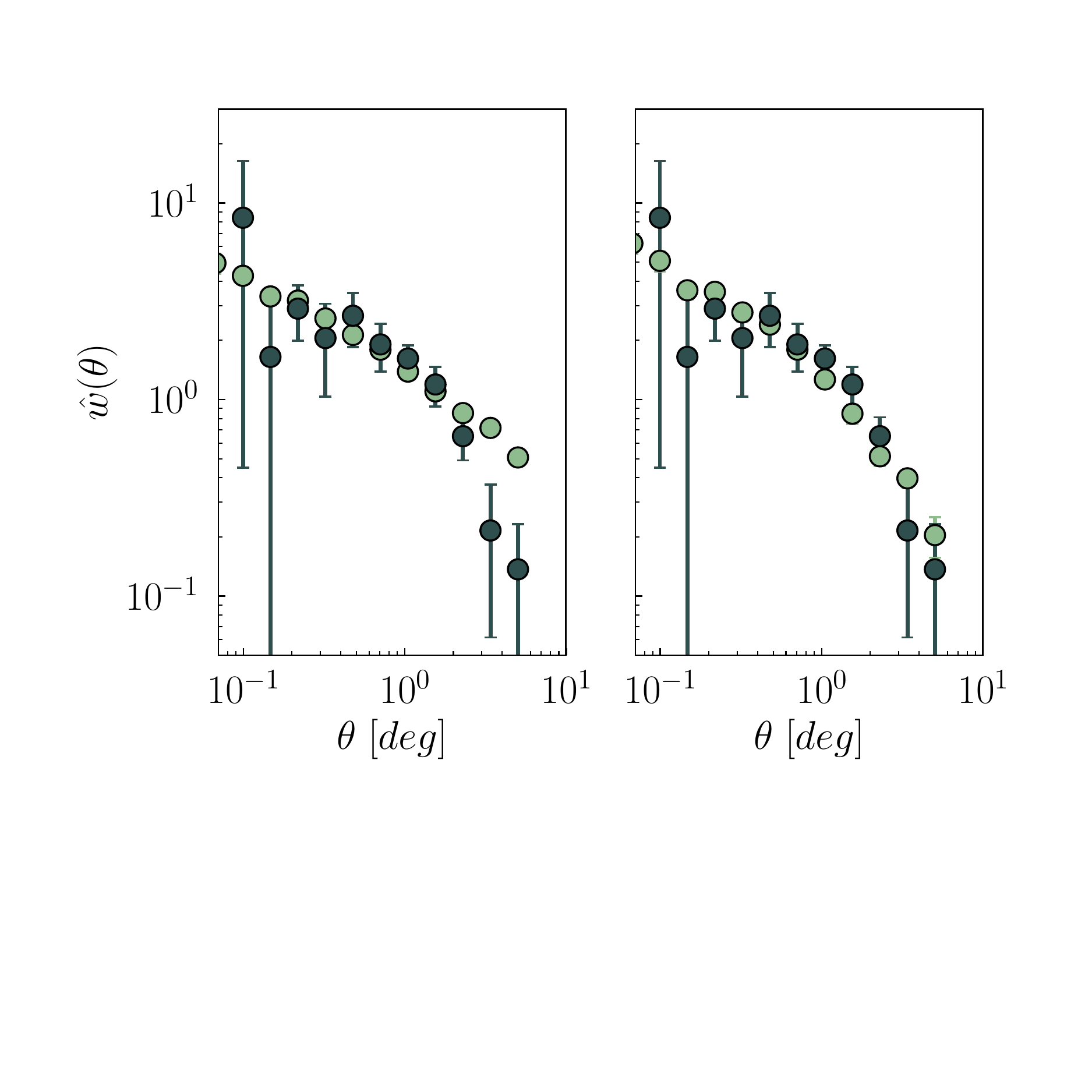}
\end{center}
\vskip -3cm
\caption{The angular correlation functions for SDSS galaxies beyond a projected radius of 3 Mpc from the Coma cluster and within a projected radius of 18 Mpc  with measured recessional velocities, $cz,$ $< 10,000$ km s$^{1}$ (lighter filled circles with no error bars, left panel),
the subset of these same galaxies that have $cz$ corresponding to the Coma velocity ($6000 < cz < 8000$ km s$^{-1}$; lighter filled circles with no error bars, right panel), and
our sample of UDGs, selected over the same projected radius range (darker filled circles with error bars, both panels). The improved agreement at larger angular separations between the UDGs and Coma SDSS galaxies (right panel) suggests, under the assumption that the UDGs have a similar 2-pt correlation function in real space, that the majority of our UDGs are at the distance of the Coma cluster.}
\label{fig:angcorr}
\end{figure}

Such correspondences, either real or accidental, between the UDGs and normal galaxies can be evaluated more quantitatively. Presuming that the UDGs share similar physical correlation behavior to that of normal galaxies, the redshift slice for which the normal galaxy angular correlation function is a best match to the angular correlation function of the UDGs is likely to contain the bulk of the UDG sample. For example, comparing the angular correlation function of all of the normal galaxies that are projected outside of the center of Coma (where there is an obvious correlation between normal galaxies and UDGs), within 18 Mpc of the Coma cluster (to limit ourselves to the area for which we have so far identified UDGs), and have $cz \le $ 10,000 km s$^{-1}$ to that of the UDG sample yields a poor match (Figure \ref{fig:angcorr}, left panel). When we instead limit
the comparison to normal galaxies within a narrow redshift strip that contains the Coma cluster and associated structures ($6000 < cz/({\rm km\ s}^{-1}) < 8000)$, we find excellent agreement, especially at large angular separation where we will be sensitive to distance effects (Figure \ref{fig:angcorr}, right panel). 
We conclude that this sample of UDGs, which was chosen to be large in angular extent, is also predominantly one of physically large UDGs because it is mostly near the distance of Coma and the associated large scale structure. This is a statement for the sample as a whole, but unfortunately says little about the location of any particular UDG candidate.
A full mathematical treatment of this approach will be presented elsewhere along the lines of that performed by \cite{rahman}, once we have a larger spectroscopic redshift sample to confirm the estimated redshift distributions.

\subsection{Size Distribution}

Given a narrow range of surface brightness within $r_e$, the velocity dispersion of a galaxy, and hence the enclosed mass, is proportional to a power of $r_e$ \citep{zar17a}. Because our survey is limited to about a $\pm$ one magnitude range in surface brightness (Figure \ref{fig:selection}), we expect the physically largest systems to be the most massive and interesting. The largest confirmed UDGs, those with existing spectroscopic redshifts, are DF08 with $r_e = 4.4$ kpc and DF44 with $r_e = 4.6$ kpc \citep{vdk15a,vdk17,kadowaki17} and there are a few others that are slightly smaller but still have $r_e > 4$ kpc \citep{kadowaki17}. So far, this appears to be the upper end of the UDG size distribution.

\begin{figure}[ht]
\begin{center}
\includegraphics[width=0.4\textwidth]{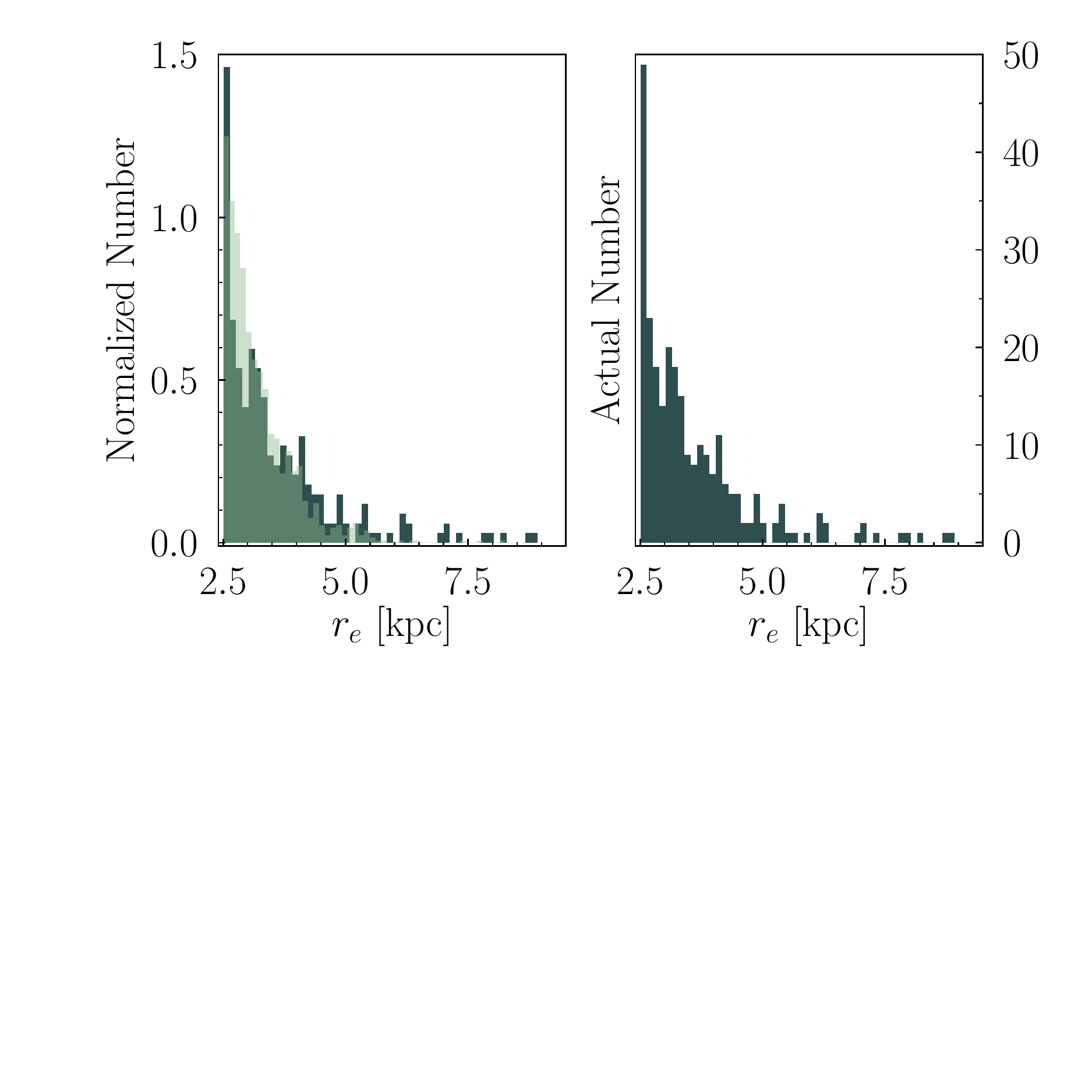}
\end{center}
\vskip -3cm
\caption{The size distribution as parameterized by the effective radius, $r_e$ for the SDSS galaxies in the survey volume (light green histogram in left panel) and the UDGs (dark green histogram in both panels). For comparison, histograms are normalized to have unit area in left panel. Galaxies with $r_e > 5$ kpc are rare in both samples. For the UDGs, we estimate $r_e$ assuming they all lie at the distance to Coma. Therefore, some of the unusually large systems may simply be much closer to us than the Coma cluster. The UDG size distribution, if most lie at the distance of Coma and its associated large scale structure, is consistent with that of SDSS galaxies.}
\label{fig:sizehist}
\end{figure}

The inferred size distribution of our UDGs (assuming they lie at the distance of Coma) is qualitatively consistent with the size distribution of SDSS galaxies in the region (Figure \ref{fig:sizehist}) and shows a similar rapid decline toward larger sizes. The distance assumption is based on the earlier  comparison of angular correlation functions, from which we concluded that the bulk of our UDG sample lies at roughly the distance of the Coma cluster. This inference is supported by the redshift distribution of the SDSS galaxies in this region (Figure \ref{fig:zplot}), where the dominant structure is a filament that includes the Coma cluster and there are several large voids in the foreground to Coma. For both UDGs and SDSS galaxies, objects with $r_e > 5$ kpc are quite rare (particularly considering galaxies with $M_g > -20$). 

\begin{figure}[ht]
\begin{center}
\includegraphics[width=0.5\textwidth]{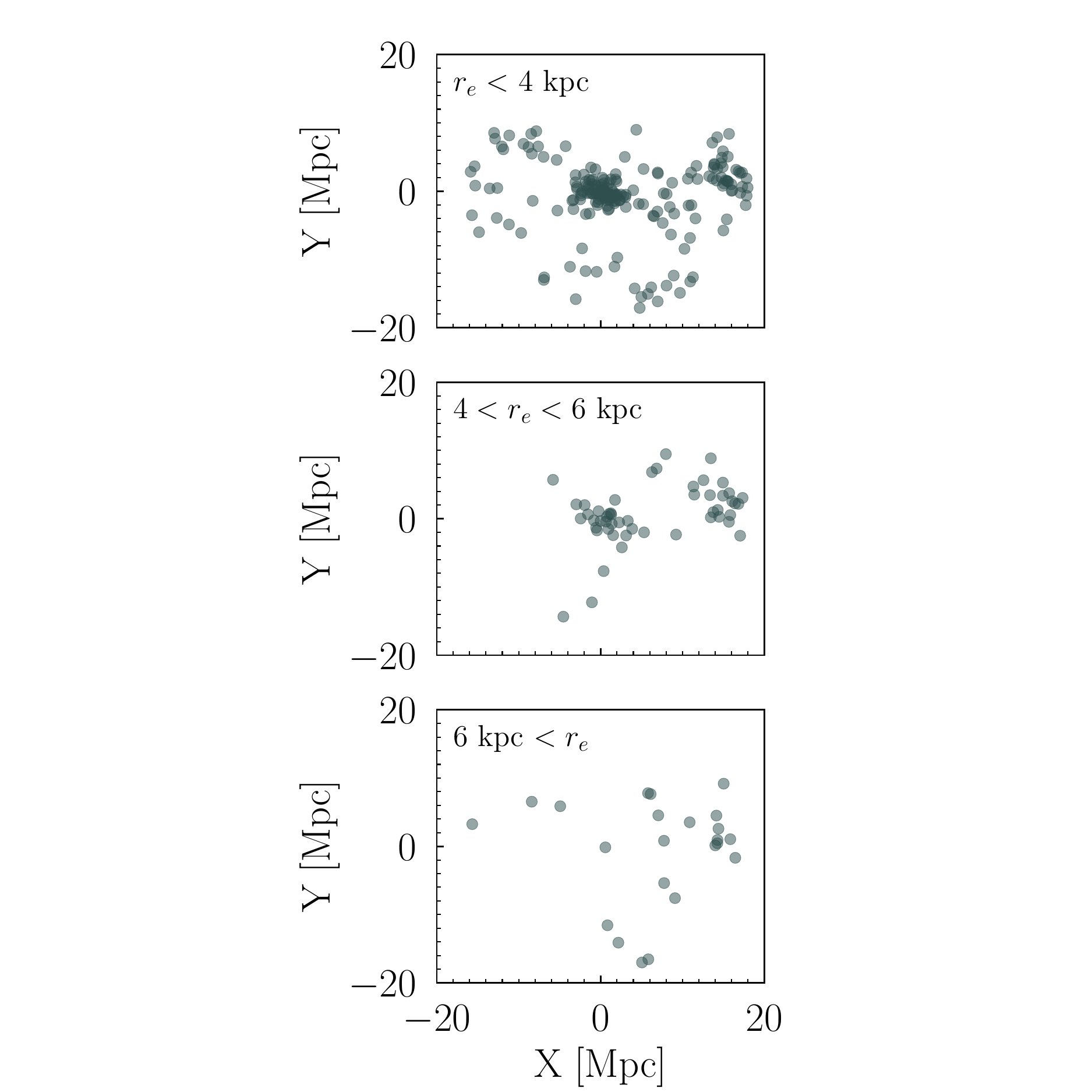}
\end{center}
\caption{The spatial distribution of UDGs of different sizes. The abundance of large UDGs ($4 < r_e/{\rm kpc} < 6 $) in Coma demonstrates that the large inferred physical sizes cannot be entirely due to erroneous adopted distances. The lack of any corresponding concentration suggests that inferred sizes $>$ 6 kpc are, at least, mostly incorrect. Orientation is the same as in Figure \ref{fig:posplot2}.}
\label{fig:size}
\end{figure}

There are a number of candidate UDGs with $r_e > 5$ kpc. We show in Figure \ref{fig:size} that even the UDGs with $4 < r_e/{\rm kpc} < 6$ congregate around the Coma cluster, demonstrating that the bulk of these are indeed at the distance of Coma, but that the UDGs with $r_e > 6$ kpc do not, suggesting that these are not physically associated with the Coma cluster and therefore not at that distance. These are likely to be in the foreground and so have smaller physical sizes than we assigned them. In fact, some congregation of these ``largest" UDGs is seen near what we have labeled feature C, which we previously noted is likely a superposition of both a local overdensity and overdensities at and beyond the Coma distance. Furthermore, some of these candidates have been spectroscopically confirmed as local in ongoing work (Kadowaki et al. in prep).
Nevertheless, there appear to be a significant number of analogs to DF 08 and 44 in size that have yet to be spectroscopically confirmed, both within Coma and in the field.

Combining a quantitative version of this analysis and a much larger sample of UDG candidates and SDSS galaxies could lead to a determination of the upper end of the UDG size distribution. The angular correlation between UDG candidates and SDSS galaxies, where we can use the redshifts of the SDSS galaxies to isolate slices in the 3-D distribution, will vanish when selecting UDGs that would be larger than those existing in reality.

\subsection{Color Distribution}

The star formation history of large UDGs is critical to establishing the physical reason for the apparent low star formation efficiency of
these systems. DF44, for example, is estimated to have a total mass close to that of the Milky Way and yet only contain 1\% as many stars \citep{vdk15b}. Has star formation ceased in all UDGs? Is it driven by environmental factors? Large samples such as that presented here, with colors, can begin to address such questions. 

\begin{figure}[ht]
\begin{center}
\includegraphics[width=0.5\textwidth]{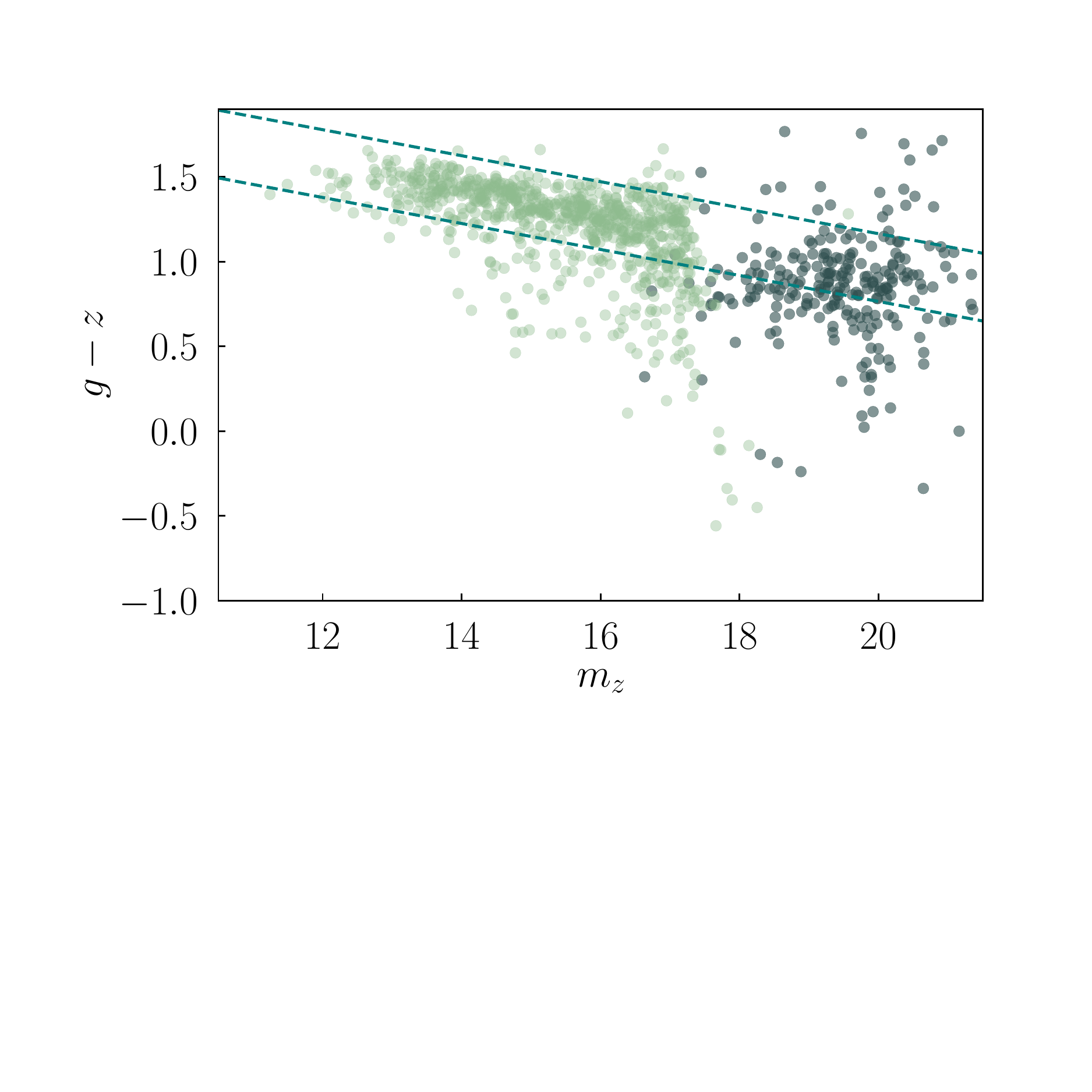}
\end{center}
\vskip -3cm
\caption{The color-apparent magnitude diagram for SDSS galaxies with $0.018 < z < 0.028$ within 2$^\circ$ of the Coma cluster (light green) and for our UDGs (dark green). Colors for the UDGs are measured as the difference in modeled central surface brightnesses. The UDGs span the range of colors of the SDSS galaxies. The dashed lines indicate the region defined by eye to follow the Coma cluster red sequence. For galaxies at the distance of the Coma cluster, the abscissa corresponds to $-24.6< M_z <-13.6$.}
\label{fig:colormag}
\end{figure}

With the data in hand we examine broad band colors, $g-z$, across the whole sample and as a function of environment. To minimize the effect of noisy measurements, we choose to define our $g-z$ measurement as the difference in corresponding central surface brightnesses. This could introduce a bias if UDGs have color gradients, but there is no measurement yet of systematic color gradients. In Figure \ref{fig:colormag} we compare the color magnitude distribution to SDSS galaxies within 2$^\circ$ of Coma that are at the distance of Coma ($0.018 < z < 0.028$). We plot color against $z$ band magnitude, which is our best proxy for stellar mass. The UDGs span a range of colors from as red as the extrapolated sloped red sequence to well into the blue cloud. One interesting aspect of the Figure is that there are only a few objects that are redder than the extrapolated red sequence. Given that one potential source of contamination is high redshift galaxy clusters \citep[cf.][]{gonzalez}, it is reassuring that we are not detecting a large population of such red objects. 

It is evident from this CMD that even large UDGs must span a range of formation scenarios given the diverse properties. These large UDGs are not, entirely, fully quenched galaxies that follow directly from the SDSS red sequence population.
Interpreting the color-magnitude diagram, CMD, is not entirely straightforward because there are both mean age and metallicity effects. 
For two galaxies with the same stellar mass and mean age, the mean stellar metallicity can be quite different depending on the star formation history and the chemical enrichment history. Spectrsoscopic analysis of UDG populations, as recently done \citep{gu,ferre,ruiz} is a more robust approach that can help constrain the range of options. Even so, CMDs are valuable because they can define samples or highlight the most interesting objects for follow-up spectroscopy.

\begin{figure}[ht]
\begin{center}
\includegraphics[width=0.5\textwidth]{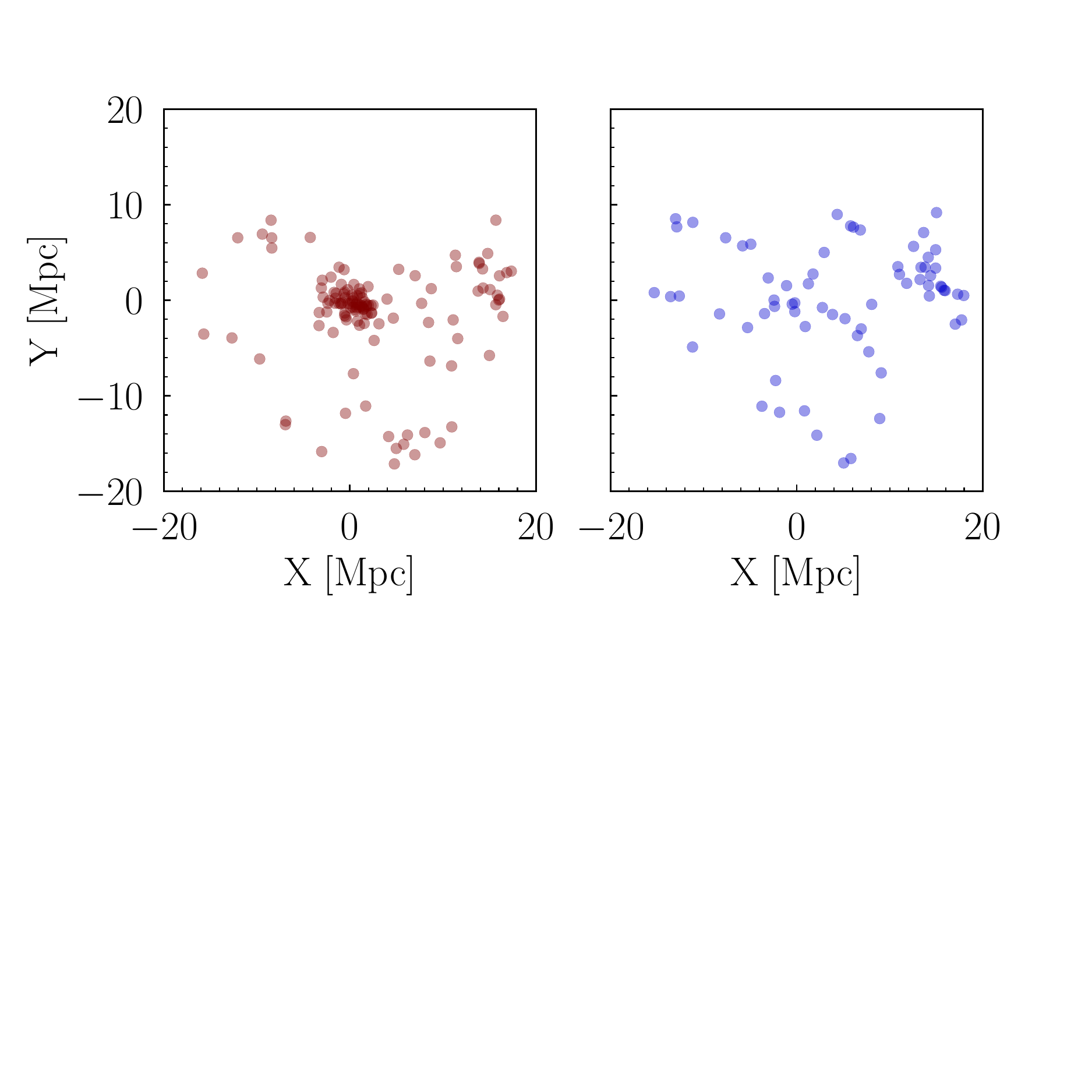}
\end{center}
\vskip -4cm
\caption{The spatial distribution of UDGs depending on $g-z$ color. The red sequence (left panel) and blue cloud (right panel) selection is described in the text. There is slight evidence for clustering near Coma for the blue galaxies and most red galaxies are found in the Coma cluster or the other structures identified in Figure \ref{fig:posplot2}.}
\label{fig:color}
\end{figure}

To investigate how the colors may differ as a function of environment over the survey region we divide the sample by color (Figure \ref{fig:color}). We select the red sequence galaxies visually and extrapolate the slope to fainter magnitudes (shown in Figure \ref{fig:colormag}). The blue galaxies are simply those that are at least 0.1 mag bluer than the blue side of the defined red sequence region.
There are two clear results in the Figure. 
First, both red and blue UDGs are found in the vicinity of the Coma cluster, although there is a preponderance of red galaxies.
Once kinematics are available for a larger number of these, it will be interesting to determine if there are indications of different dynamical histories for UDGs split by color. Second, feature C is proportionally more evident among the blue UDGs than the other overdensities. To some degree this can be an artifact of small numbers, but it may also reflect the higher level of contamination by nearby, smaller low surface brightness galaxies that we have discussed before.  Distinguishing whether this is a physical result or due to contamination will be straightforward with statistics from a larger survey.

\section{Plans for the Full SMUDGES Program}

The current analysis explores roughly 300 deg$^2$. The completed DESI imaging footprint is envisioned to extend to 14,000 deg$^2$. Even assuming that some sizable portion is not usable for our purposes due to extreme IR cirrus contamination, we may be facing something in the neighborhood of 10,000 deg$^2$ of high quality imaging, about 33 times more than that which we present here. Such sky coverage would suggest that we will identify a final sample of thousands of UDGs of large angular extent. This, in turn, helps highlight some daunting challenges in context. 

The original processing run of the data presented here required about 30 days on a single desktop-quality machine and utilized more than 3 TB of storage suggesting that the full analysis would require well over three years of CPU time and more than 100 TB of storage.  This is a significant, but not overwhelming challenge. However, one is also likely to require many runs with artificial sources to determine completeness corrections. That work will be the subject of a subsequent study in this series, and one may, for a variety of reasons, want to repeat at least parts of the original data processing. Now we have ported the pipeline to a high performance computer (HPC) on which we can process data at a rate of $\sim$ 1 hour/exposure. There are typically $\approx$ 3 exposures/deg$^2$, so processing the nearly 10,000 sq. degrees of the full survey will require $\approx$ 30,000 hours.  We expect to use $\sim$ 30 cores on the HPC so approximately six weeks of computer time is required. The enterprise will take somewhat longer than this because there will undoubtedly be instances where human intervention is needed to address a problem. Nevertheless, this timescale is eminently practical. Regarding the disk space requirements, we will compress our data by $\sim$6 and only save processed images. This approach suggests a requirement of $\sim$ 15 TB.

The data processing and UDG identification is comprised of several steps, some of which one might want to revisit and improve. As such, we will save intermediate steps. For example, it is possible that a new object identification procedure, either in the original source selection or the subsequent ``morphological'' classification, will be developed and one would simply want to run that on the cleaned images. These intermediate steps are not presented with this paper, but our intent is to make them easily available for the full survey.

The second, and much more serious challenge, is the lack of distance information for the UDG candidates. 
With an initial outlay of 4 nights of spectroscopic time on the Large Binocular Telescope (LBT), \cite{kadowaki17} obtained redshifts for 5 UDGs. At this rate we would require over 200 LBT nights just to observe the UDGs identified in this study, not the full sample we expect. We have also begun a program to target UDGs at 21cm wavelengths because we expect, on the basis of the significant fraction with blue colors (Figure \ref{fig:colormag}) and recent {\small HI} results \citep{spekkens}, that at least some UDGs have significant gas reservoirs. Although we do not yet have results to report with regards to the sample presented here, similar observations \citep{spekkens} suggest that such observations can provide redshifts, although only moderately more efficient in terms of large telescope time. Even so, these observations have great potential to also provide a measurement of the internal kinematics and a direct measure of the gas content of UDGs, both of which will be critical in understanding the physical nature of UDGs. Nevertheless, we cannot expect to have spectroscopic redshifts for the bulk of our final UDG sample for the foreseeable future.

\section{Summary}

We present our first batch processing of DESI pre-imaging data of a region around the Coma galaxy cluster in search of large ($r_e \gtrsim 5.3$\arcsec\ or equivalently $\ge 2.5$ kpc for those systems at the distance of the Coma cluster) ultra-diffuse galaxies (UDGs). We have termed this program as an effort that is Systematically Measuring Ultra-Diffuse Galaxies (SMUDGes). Due to the large data volume, a search through the entire DESI pre-imaging data must be automated. We have developed a complete pipeline that 1) queries the data repository, 2) retrieves the data,  3) performs some additional basic data reduction to prepare that data, 4) subtracts all high surface brightness objects and corrects for extended emission from the brightest of those, 5) wavelet transforms the residual image, 6) identifies significant peaks of emission in a series of those transformed images, 7) compares results from various photometric bands, 8) rejects defects and distinguishes among real sources of emission using machine learning methods, and 9)  fits simple models to the stacked image of UDG candidates to do the final selection for large UDGs and provide their basic parameters. The pipeline is working on a high performance cluster and is ready to continue processing the full survey.

We illustrate the results obtained with this pipeline by presenting a catalog of UDGs drawn from an area covering $\approx$ 290 sq. degree centered on the Coma galaxy cluster. By centering on Coma we are able to compare our detections directly to those from two recent surveys for UDGs \citep{vdk15a,yagi}. Those surveys cover a significantly smaller area, but all cover the core of the Coma cluster. From a comparison of results, we conclude that we are achieving comparable sensitivity and that where there are distinctions between our catalog and the literature, the differences arise from details of the object classification rather than the detections themselves. 

We present a catalog of 275 candidate large UDGs. Using the catalog, we show that the majority of our UDGs are likely to be at roughly the distance of Coma based both on the number of galaxies that congregate around Coma and on the details of the angular correlation function. Accepting this result, we  conclude that our sample contains a significant number of analogs of DF 44 in size, where the effective radius exceeds 4 kpc. The conclusion is further confirmed by the angular clustering of these apparently large UDGs at the location of the Coma cluster. In contrast, systems whose effective radius would exceed 6 kpc do not cluster near Coma and we  conclude that for these systems we are overestimating the distance and the size by incorrectly associating them with Coma. 

We find that the $g-z$ color of UDGs spans the full range from the red sequence to the blue edge of the blue cloud. As such, even the large UDGs are likely to reflect a range of formation histories. However, the majority have colors that are consistent with being low stellar mass versions of red sequence galaxies. Additional evidence for a range of processes being at play in UDG formation is that we find an overabundance of blue UDG candidates in the lower density enhancements and the field. Redshifts are necessary to confirm that these are physically large galaxies.

The statistical treatment of large UDG samples will enable us to circumvent the critical challenge that we will be lacking spectroscopic redshifts for the majority of this sample. As such, we will benefit from the largest possible samples. Our analysis of the Legacy Surveys in search of UDGs, which we refer to as SMUDGes, promises to be the best such source until the advent of analogous analysis of the deeper LSST imaging. 

\acknowledgments

DZ, RD, JK, and HZ acknowledge financial support from NASA ADAP NNX12AE27G and NSF AST-1713841. AD's research is supported by the National Optical Astronomy Observatory.
DMD acknowledges support by Sonderforschungsbereich (SFB) 881
``The Milky Way System'' of the German Research Foundation (DFG), particularly through sub-project A2.
An allocation of computer time from the UA Research Computing High Performance Computing (HPC) at the University of Arizona and the prompt assistance of the associated computer support group is gratefully acknowledged.
 
This research has made use of the NASA/IPAC Extragalactic Database (NED), which is operated by the Jet Propulsion Laboratory, California Institute of Technology, under contract with NASA. 

The Legacy Surveys consist of three individual and complementary projects: the Dark Energy Camera Legacy Survey (DECaLS; NOAO Proposal ID \# 2014B-0404; PIs: David Schlegel and Arjun Dey), the Beijing-Arizona Sky Survey (BASS; NOAO Proposal ID \# 2015A-0801; PIs: Zhou Xu and Xiaohui Fan), and the Mayall $z$-band Legacy Survey (MzLS; NOAO Proposal ID \# 2016A-0453; PI: Arjun Dey). DECaLS, BASS and MzLS together include data obtained, respectively, at the Blanco telescope, Cerro Tololo Inter-American Observatory, National Optical Astronomy Observatory (NOAO); the Bok telescope, Steward Observatory, University of Arizona; and the Mayall telescope, Kitt Peak National Observatory, NOAO. The Legacy Surveys project is honored to be permitted to conduct astronomical research on Iolkam Du'ag (Kitt Peak), a mountain with particular significance to the Tohono O'odham Nation.

NOAO is operated by the Association of Universities for Research in Astronomy (AURA) under a cooperative agreement with the National Science Foundation.

This project used data obtained with the Dark Energy Camera (DECam), which was constructed by the Dark Energy Survey (DES) collaboration. Funding for the DES Projects has been provided by the U.S. Department of Energy, the U.S. National Science Foundation, the Ministry of Science and Education of Spain, the Science and Technology Facilities Council of the United Kingdom, the Higher Education Funding Council for England, the National Center for Supercomputing Applications at the University of Illinois at Urbana-Champaign, the Kavli Institute of Cosmological Physics at the University of Chicago, Center for Cosmology and Astro-Particle Physics at the Ohio State University, the Mitchell Institute for Fundamental Physics and Astronomy at Texas A\&M University, Financiadora de Estudos e Projetos, Fundacao Carlos Chagas Filho de Amparo, Financiadora de Estudos e Projetos, Fundacao Carlos Chagas Filho de Amparo a Pesquisa do Estado do Rio de Janeiro, Conselho Nacional de Desenvolvimento Cientifico e Tecnologico and the Ministerio da Ciencia, Tecnologia e Inovacao, the Deutsche Forschungsgemeinschaft and the Collaborating Institutions in the Dark Energy Survey. The Collaborating Institutions are Argonne National Laboratory, the University of California at Santa Cruz, the University of Cambridge, Centro de Investigaciones Energeticas, Medioambientales y Tecnologicas-Madrid, the University of Chicago, University College London, the DES-Brazil Consortium, the University of Edinburgh, the Eidgenossische Technische Hochschule (ETH) Zurich, Fermi National Accelerator Laboratory, the University of Illinois at Urbana-Champaign, the Institut de Ciencies de l'Espai (IEEC/CSIC), the Institut de Fisica d'Altes Energies, Lawrence Berkeley National Laboratory, the Ludwig-Maximilians Universitat Munchen and the associated Excellence Cluster Universe, the University of Michigan, the National Optical Astronomy Observatory, the University of Nottingham, the Ohio State University, the University of Pennsylvania, the University of Portsmouth, SLAC National Accelerator Laboratory, Stanford University, the University of Sussex, and Texas A\&M University.

BASS is a key project of the Telescope Access Program (TAP), which has been funded by the National Astronomical Observatories of China, the Chinese Academy of Sciences (the Strategic Priority Research Program ``The Emergence of Cosmological Structures" Grant \# XDB09000000), and the Special Fund for Astronomy from the Ministry of Finance. The BASS is also supported by the External Cooperation Program of Chinese Academy of Sciences (Grant \# 114A11KYSB20160057), and Chinese National Natural Science Foundation (Grant \# 11433005).

The Legacy Survey team makes use of data products from the Near-Earth Object Wide-field Infrared Survey Explorer (NEOWISE), which is a project of the Jet Propulsion Laboratory/California Institute of Technology. NEOWISE is funded by the National Aeronautics and Space Administration.

The Legacy Surveys imaging of the DESI footprint is supported by the Director, Office of Science, Office of High Energy Physics of the U.S. Department of Energy under Contract No. DE-AC02-05CH1123, by the National Energy Research Scientific Computing Center, a DOE Office of Science User Facility under the same contract; and by the U.S. National Science Foundation, Division of Astronomical Sciences under Contract No. AST-0950945 to NOAO.

\facilities{Blanco}

\software {galfit \citep{peng}, keras \citep{keras},  lmfit \citep{newville}, scipy \citep{jones}, sep \citep{sep},  sextractor \citep{bertin} }

\clearpage

\begin{deluxetable}{@{\extracolsep{6pt}}lrrrrrrrrrrrrrrrr@{}}
\setlength\tabcolsep{2 pt}
\rotate
\tablecaption{Ultra Diffuse Galaxy Properties}

\tablehead{
\multicolumn{3}{c}{}&
\multicolumn{6}{c}{\textbf{S\'ersic index fixed at $n$ = 1}}&  
\multicolumn{4}{c}{\textbf{S\'ersic index, $n$, floating}} \\
\cmidrule(lr){4-9} \cmidrule(lr){10-13}\\
\colhead{Name}&
\colhead{$\alpha$}&
\colhead{$\delta$}&
\colhead{$\mu_{0,g}$}&
\colhead{$\mu_{0,r}$}&
\colhead{$\mu_{0,z}$}&
\colhead{r$_{e}$}&
\colhead{$b/a$}&
\colhead{$\theta$}&
\colhead{m$_{g}$}&
\colhead{m$_{r}$}&
\colhead{m$_{z}$}&\\
&
&
&
&
&
&
[arcsec]
&
&
&
\\
}
\startdata
SMDG1212080+281630&183.033401&28.275014&24.4$\pm$0.4&24.7$\pm$0.5&24.8$\pm$0.9&5.6$\pm$0.6&0.88$\pm$0.08&87$\pm$28&21.1$\pm$0.1&21.1$\pm$0.1&20.6$\pm$0.2\\
SMDG1212084+290349&183.035145&29.063514&...&24.2$\pm$0.2&23.6$\pm$0.5&6.1$\pm$0.5&0.67$\pm$0.05&156$\pm$7&...&...&19.6$\pm$0.3\\
SMDG1212454+273507&183.189350&27.585352&24.7$\pm$0.4&24.6$\pm$0.4&23.6$\pm$0.8&5.4$\pm$0.6&0.59$\pm$0.05&140$\pm$7&21.4$\pm$0.1&20.8$\pm$0.2&20.1$\pm$1.2\\
SMDG1213061+294549&183.275309&29.763710&24.0$\pm$0.1&23.6$\pm$0.1&23.0$\pm$0.2&10.6$\pm$0.2&0.79$\pm$0.01&116$\pm$3&18.3$\pm$0.1&18.0$\pm$0.1&17.7$\pm$0.1\\
SMDG1213219+293401&183.341232&29.566810&25.4$\pm$0.4&25.1$\pm$0.4&...&6.4$\pm$0.8&0.52$\pm$0.05&76$\pm$6&21.9$\pm$0.1&21.0$\pm$0.2&...\\
SMDG1213235+264641&183.347943&26.777990&25.1$\pm$0.3&24.6$\pm$0.3&24.4$\pm$0.6&6.4$\pm$0.5&0.52$\pm$0.03&80$\pm$4&21.2$\pm$0.1&20.9$\pm$0.1&...\\
SMDG1213512+282109&183.463343&28.352505&24.2$\pm$0.2&24.0$\pm$0.2&23.7$\pm$0.4&5.5$\pm$0.3&0.83$\pm$0.04&41$\pm$10&20.0$\pm$0.1&19.8$\pm$0.1&18.6$\pm$1.2\\
SMDG1214010+293203&183.504356&29.534183&25.7$\pm$0.4&25.5$\pm$0.4&...&6.3$\pm$0.9&0.72$\pm$0.09&106$\pm$16&21.8$\pm$0.1&21.3$\pm$0.3&...\\
SMDG1214279+294034&183.616139&29.675991&25.2$\pm$0.2&25.2$\pm$0.4&24.3$\pm$0.6&6.1$\pm$0.5&0.70$\pm$0.05&99$\pm$8&21.6$\pm$0.1&20.9$\pm$0.1&20.6$\pm$0.2\\
SMDG1214419+274955&183.674463&27.831858&24.6$\pm$0.2&24.1$\pm$0.1&23.8$\pm$0.3&8.4$\pm$0.4&0.97$\pm$0.04&91$\pm$43&19.4$\pm$0.1&18.9$\pm$0.1&19.0$\pm$0.1\\
\enddata

\label{tab:catalog}

\tablecomments{Table \ref{tab:catalog} is published in its entirety in the machine-readable format. A portion is shown here for guidance regarding its form and content.}
\end{deluxetable}

\clearpage

\begin{deluxetable}{ll|ll}
\tabletypesize{\footnotesize}
\tablecaption{Crosslisting Large Dragonfly Coma UDGs with SMUDGES}
\tablewidth{0pt}
\tablehead{
\colhead{Dragonfly ID}&
\colhead{SMDG ID}&
\colhead{Dragonfly ID}&
\colhead{SMDG ID}\\
\\
}
\startdata
DF01&SMDG1259142+290717  &DF28&SMDG1259304+274450  \\
DF03&SMDG1302166+285717  &DF29&SMDG1258050+274358  \\
DF04&SMDG1302334+283452  &DF30&SMDG1253151+274115  \\
DF06&SMDG1256297+282640  &DF31&SMDG1255062+273727  \\
DF07&SMDG1257017+282325  &DF32&SMDG1256284+273706  \\
DF08&SMDG1301304+282228  &DF34&SMDG1256129+273250  \\
DF09&SMDG1256228+281955  &DF35 &...           \\
DF12 &...           &     DF36&SMDG1255554+272736  \\
DF14 &...           &     DF39&SMDG1258104+271911  \\
DF15&SMDG1258164+275330  &DF40&SMDG1258011+271126  \\
DF17&SMDG1301582+275011  &DF41&SMDG1257190+270556  \\
DF18 &...           &     DF42&SMDG1301191+270315  \\
DF19&SMDG1304052+274804  &DF44&SMDG1300580+265835  \\
DF25&SMDG1259487+274639  &DF46&SMDG1300473+264700  \\
DF26&SMDG1300206+274712  &DF47&SMDG1255481+263352  \\
\enddata

\label{tab:DF}

\end{deluxetable}

\clearpage 

\begin{deluxetable}{lrl|lrl}
\tablecaption{Crosslisting Large Yagi Coma UDGs with SMUDGES}
\tablewidth{0pt}
\tablehead{
\colhead{Koda ID}&
\colhead{Code}&
\colhead{SMDG ID}&
\colhead{Koda ID}&
\colhead{Code}&
\colhead{SMDG ID}
\\
}
\startdata
11 &&SMDG1300580+265835&       443 &B&SMDG1258403+283905         \\
13 &&SMDG1301158+271238&       486 &&SMDG1259142+290717           \\
14 &&SMDG1301191+270315&       494 &&SMDG1256514+271338           \\
16 &B/S&SMDG1301224+264950&    501 &&SMDG1257190+270556           \\
22 &S&SMDG1302148+270843&      507 &&SMDG1258011+271126           \\
37 &C&SMDG1300356+272951&      526 &B&SMDG1256489+273650          \\
53 &&SMDG1301124+274525&       553 &&SMDG1257243+274343           \\
92 &B&SMDG1300217+281341&      571 &&SMDG1257534+273202           \\
93 &&SMDG1300206+274712&       577 &&SMDG1258050+274358           \\
98 &S&SMDG1300231+274820&      581 &&SMDG1258104+271911           \\
165 &&SMDG1301582+275011&      583 &&SMDG1258132+272522           \\
166 &B&SMDG1301584+275454&     584 &&SMDG1258145+272429           \\
168 &S&SMDG1302013+280508&     629 &S&SMDG1257346+275440          \\
178 &&SMDG1300279+283730&      641 &&SMDG1257515+274923           \\
194 &&SMDG1301304+282228&      649 &S&SMDG1258033+280805          \\
212 &S&SMDG1302013+291222&     653 &S&SMDG1258078+275444          \\
215 &&SMDG1302166+285717&      654 &S&SMDG1258067+280112          \\
230 &B&SMDG1259174+270140&     660 &&SMDG1258164+275330           \\
215 &&SMDG1302166+285717&      654 &S&SMDG1258067+280112          \\
230 &B&SMDG1259174+270140&     660 &&SMDG1258164+275330           \\
281 &&SMDG1259413+273148&      680 &&SMDG1257017+282325           \\
285 &&SMDG1259487+274639&      695 &C&SMDG1257470+284643          \\
320 &&SMDG1258472+280724&      698 &C&SMDG1257506+283756          \\
328 &B&SMDG1258495+274216&     739 &&SMDG1255062+273727           \\
348 &&SMDG1259088+275736&      743 &S&SMDG1255201+273027          \\
352 &&SMDG1259106+275415&      774 &&SMDG1256129+273250           \\
390 &S&SMDG1259503+281123&     782 &&SMDG1256284+273706           \\
407 &S&SMDG1300054+275333&     787 &&SMDG1256391+274055           \\
416 &A&SMDG1300092+280829&     819 &B&SMDG1255584+280357          \\
436 &S&SMDG1300292+275924&     851 &&SMDG1256228+281955            
\enddata
\label{tab:koda}
\tablenotes{
The large \cite{yagi} UDG candidates that do not have counterparts in our detection catalog, prior to cuts made on size, brightness, and morphology are 4, 94, 199, 218, 231, 240, 257, 275, 310, 370, 425, 453, 456, 471, and 569. The Code column presents our reason for excluding the candidate from our final UDG catalog (B = too bright, S = too small, A = too elongated, C = classified as other type of source). 
\hfill}
\end{deluxetable}

\appendix
\section[]{Deep Learning Classifier}
\label{app:classifier} 
We selected the DenseNet-201 model for our initial testing because their results suggest that it might provide better accuracy than other recent algorithms when  trained on relatively small datasets \citep{huang} . The Keras version of this application is designed to classify 1000 categories and requires modification to accommodate a binary decision (UDG or not UDG).  Therefore, the softmax final layer is replaced with a dense layer with a sigmoid activation as a single output. This is preceded by a global average pooling layer to reduce overfitting \citep{lin}.  The Keras implementation of this model also eliminated the dropout layers \citep{sri} that were part of the of the original version \citep{huang}.  Because of our small dataset, we added this function back into the transition layers to further reduce overfitting.  All layers of the model are allowed to vary during training.  The base layers are initially assigned weights derived from the Imagenet dataset \citep{deng} while weights of the modified top layers are randomly assigned.  

Our 201 $\times$ 201 pixel thumbnails are resized to 150 $\times$ 150 and then divided into a train/validation set (80\%) and a separate test set (20\%).  The test set is not evaluated until all hyperparameters are finalized. The train/validation set was further divided into four folds to be used for cross-validation.  Two versions are created for each fold.  One version contains the original data required for validation and the other is augmented and used for training.  Augmentation is needed to expand our limited dataset and consists of random flipping, rotation and translation of the original images. To prevent excessive padding of image edges, rotation is limited to 10$^\circ$ and translation is limited to a distance of 7 pixels from the center after random flipping.  The original dataset is unbalanced with an excess of non-UDGs and, therefore, UDGs are augmented with 6 transformations while others are augmented with three.  Cross-validation is performed for each fold with the training set composed of the augmented versions of the other three folds and testing done with the unaugmented version.  For optimization, we use Adam \citep{king} with beta\_1 = 0.9, beta\_2 = 0.999, epsilon = 0.0000001, and decay = 0.0.  The dataset is trained for 100 epochs with an initial learning rate of 0.00005.  The learning rate is decreased by multiplying by 0.3 every 25 epochs. The Keras implementation of binary crossentropy is used as the loss function.  We investigated dropout fractions of 0.2, 0.3, and 0.4 and found that a value of 0.3 gave the best results during cross-validation.  After finalizing hyperparameters using cross-validation, the four augmented folds are combined into a single training set and run on the test set to provide our final results as described in Section \ref{Automated Classification}.

\end{document}